\documentclass[journal]{IEEEtran}

\pdfoutput=1

\usepackage{cite}

\ifCLASSINFOpdf
\else
\fi

\usepackage{url}
\usepackage[font=footnotesize]{subfig}
\usepackage{graphicx}

\usepackage{amsmath}
\usepackage{algorithm}
\usepackage{algorithmicx}
\usepackage{algpseudocode}
\usepackage{url}
\usepackage{multirow}
\usepackage{array}
\usepackage{longtable}
\usepackage{graphicx}

\usepackage{color}
\newcommand*{\old}{}
\newcommand*{\new}{}

\begin{document}

\title{Analyzing Interfaces and Workflows \\ for Light Field Editing}

\author{Marta~Ortin, Adrian~Jarabo, Belen~Masia, Diego~Gutierrez\\
 Universidad de Zaragoza - I3A, Spain
}
\markboth{Ortin \MakeLowercase{\textit{et al.}}: Analyzing Interfaces and Workflows for Light Field Editing}{}

\makeatletter
\def\ps@IEEEtitlepagestyle{
  \def\@oddfoot{\mycopyrightnotice}
  \def\@evenfoot{}
}
\def\mycopyrightnotice{
  {\footnotesize
  \begin{minipage}{\textwidth}
  Copyright~\copyright~2017 IEEE. Personal use of this material is permitted. Permission from IEEE must be obtained for all other uses, in any current or future media, including reprinting/republishing this material for advertising or promotional purposes, creating new collective works, for resale or redistribution to servers or lists, or reuse of any copyrighted component of this work in other works. DOI: 10.1109/JSTSP.2017.2746263
  \end{minipage}
  }
}

\maketitle

\begin{abstract}

With the increasing number of available consumer light field cameras, such as Lytro$^{\text{TM}}$, Raytrix$^{\text{TM}}$, or Pelican Imaging$^{\text{TM}}$, this new form of photography is progressively becoming more common. However, there are still very few tools for light field editing, and the interfaces to create those edits remain largely unexplored. Given the extended dimensionality of light field data, it is not clear what the most intuitive interfaces and optimal workflows are, in contrast with well-studied 2D image manipulation software. In this work we provide a detailed description of subjects' performance and preferences for a number of simple editing tasks, which form the basis for more complex operations. We perform a detailed state sequence analysis and hidden Markov chain analysis based on the sequence of tools and interaction paradigms users employ while editing light fields. These insights can aid researchers and designers in creating new light field editing tools and interfaces, thus helping to close the gap between 4D and 2D image editing.
\end{abstract}


\small{\textbf{IEEE Reference:} Marta Ortin, Adrian Jarabo, Belen Masia, and Diego Gutierrez. Analyzing Interfaces and Workflows for Light Field Editing. IEEE Journal of Selected Topics in Signal Processing, vol. 11, no. 7, pp. 1162-1172, Oct. 2017. doi: 10.1109/JSTSP.2017.2746263}

\section{Introduction}
\label{section:introduction}
Light fields are four-dimensional representations of a scene, where the two extra dimensions code angular information about the scene being captured. Leveraging this additional wealth of information allows to create effects such as small parallax shifts, synthetic refocusing after capture, or even scene reconstruction (see~\cite{Wu2017} for a recent survey on light field imaging and its applications). With the introduction of light field cameras in the consumer market (such as Lytro$^{\text{TM}}$~\cite{LYTRO}, Raytrix$^{\text{TM}}$~\cite{RAYTRIX} or the Pelican$^{\text{TM}}$ camera for mobile devices~\cite{PELICAN}), they are becoming an increasingly popular alternative to traditional 2D images.
\\
Editing traditional 2D photographs is a well-understood process with an established workflow. Moreover, existing image editing programs share the main ideas around which their interfaces are built, such as working with layers, creating masks, or basic point-and-click operations. \new{However, finding a user-friendly interface to edit light fields remains an open problem:
it is not clear what the best way to edit the four-dimensional information stored in a light field is, both algorithmically, and from a user's perspective. Our objective is to examine how subjects perform precise edit placement. This is a key issue in user-assisted processing tools for light fields~\cite{Jarabo:SIACG2011,Ao2015,Williem2016}, since processing tools often require user input of some kind; past light field processing methods have employed different techniques to do this, but it is unclear what the advantages and drawbacks of the different techniques are.} Jarabo et al.~\cite{Jarabo:SIG14} recently proposed the first comprehensive study on the topic: They evaluated a set of basic tools on the two most common light field interface paradigms: parallax-based, and focus-based. In their work, they asked participants to perform several edits on synthetic and real light fields, such as changing the color of an object, painting on its surface, or altering exposure (data was made publicly available by the authors\footnote{http://giga.cps.unizar.es/~ajarabo/pubs/lfeiSIG14/}). \new{As a result of the analysis, many valuable insights were provided on aspects such as the suitability of the provided interfaces and the irrelevance of inaccuracies in the depth information used, which is key to assert the appropriateness of using depth maps reconstructed from captured data.}

\begin{figure} [tbp]
   \centering
   \includegraphics[width=\columnwidth]{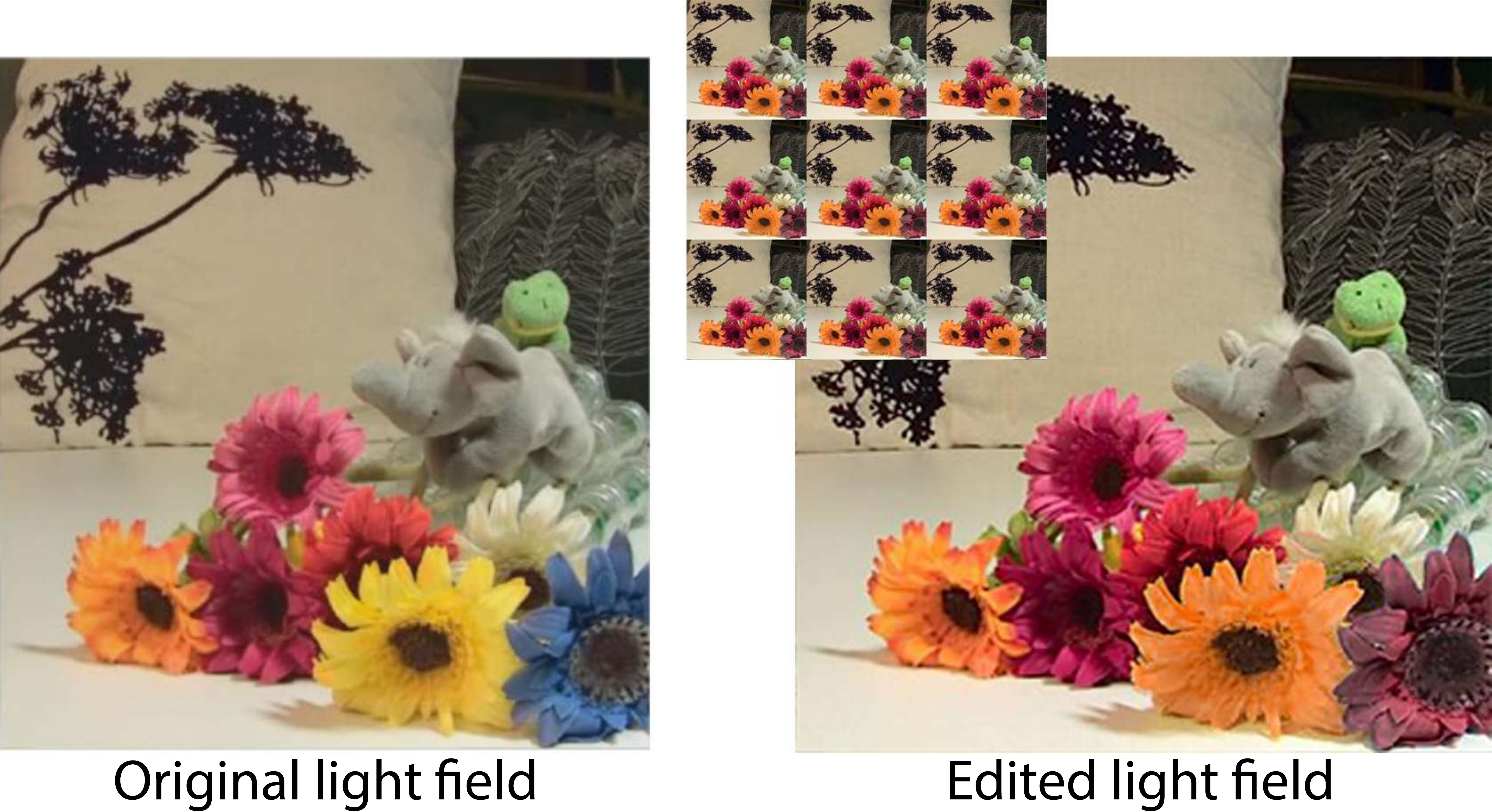}
  \caption{Example result of a real light field captured with the Lytro$^{\text{TM}}$ camera and edited with our tool. In this work we evaluate the benefits of different light field
interaction paradigms and tools, and draw conclusions to help guide future interface designs for light field editing.}
\label{figure:lightfield_example}
\end{figure}

\new{This paper presents a complementary analysis that focuses on another, but nevertheless important, aspect of editing: we analyze the subjects' preferred \textit{workflows}. We examine the temporal domain of the editing process by performing detailed state sequence analysis and Markov chain analysis, including several metrics: mean time using each editing tool, tool usage distribution, representative tool sequences, effective tool transitions, interface usage distribution, and effective interface transitions. For a number of typical scenarios (editing of surfaces, editing in free space, occlusion handling, and editing of complex geometries) we describe in detail the trends found, confirm the conclusions drawn in the previous work, and come up with novel insights.} 

\new{This paper is an extended version of our previous publication on this topic~\cite{Masia2014lf}. Our main contribution is the methodology of the analysis we perform to assess subjects' editing workflows. Analyzing workflows is a challenging task due to the large inter- and intra-user variability, given the immense number of possible editing paths one may take. We model editing workflows as sequences of states, where states can either be the tools or the interfaces used, and apply state sequence analysis and Markov modeling. Our analysis is consistent with previous findings~\cite{Jarabo:SIG14,Masia2014lf}, and offers additional insights, including:}
\begin{itemize}
\item \new{depth information is largely used for editing when available; thus, interfaces with depth are preferred;}
\item \new{there is no clear or preferred order in the use of tools for any interface;}
\item \new{despite this, the workflow is a constant iteration of drawing/erasing and checking results, and only one or two representative sequences are required to model the behaviour of 35\% to 90\% of the subjects in most cases, indicating a high agreement among them;}
\item \new{the preferred form of visualization of the light field is that based on parallax, i.e., switching between neighboring views;}
\item \new{the preferred form for editing the light field is highly dependent on the task being performed.}
\end{itemize}
This study provides a more comprehensive description of subjects' choices and preferences with regard to light field editing interfaces, together with a better understanding of the effort necessary to perform the different editing tasks. This combined knowledge is a valuable source of information for developing novel, user-friendly interfaces, as well as efficient and intuitive editing tools. 

\section{Related Work}
\label{section:related_work}
\paragraph{Editing tools} We focus here on light field editing, and refer the reader to the excellent survey by Schmidt et al.~\cite{Schmidt2014} for a more general discussion on visual content editing. Seitz and Kutulakos~\cite{Seitz2002} performed edits such as painting and scissoring by creating a voxel-based representation of the light field to propagate the edits. Zhang et al.~\cite{Zhang:2002:FLF:566654.566602} and Wang et al.~\cite{1359729} performed morphing of two light fields, for which users need to indicate corresponding polygons between them to guide the process. User interaction was also required in the Pop-Up Light Field work by Shum et al.~\cite{Shum:2004:PLF:990002.990005}, in which users marked the silhouettes of objects in multiple views, which were then used to separate the light field into depth layers. 
\new{Following the same idea, LightShop is a modular system to manipulate, composite, and render light fields~\cite{Horn2007}. Tompkin et al.~\cite{Tompkin2015} present an interaction system for 4D light field displays based on capturing the position and orientation of a pen. Nguyen et al.~\cite{Nguyen2013} edited the local light field at surfaces for material appearance editing in complex three-dimensional scenes.}  
Jarabo et al.~\cite{Jarabo:SIACG2011} devised a system for propagation of sparse edits between views, based on pixel affinity measured in the multidimensional light field space. A similar approach was later presented by Williem et al.~\cite{Williem2016}. Recently, a combined patch-based plus propagation approach was developed by Zhang and colleagues~\cite{Zhang2016}, capable of handling occlusions in tasks like object resizing or repositioning. Finally, Garces et al.~\cite{Garces2016intrinsic} proposed a method to decompose a light field into its intrinsic components, allowing users to edit the albedo and lighting independently. 
All these works focus on the particular editing tools; in contrast, we concern ourselves with the interaction required by the user, involving both the interface used, and the workflows followed by the users.
\\

\paragraph{Interfaces and workflows} 
Many different interfaces and workflows have recently been explored for a wide range of applications, including authoring of 3D models~\cite{Chen2014}, freeform design of objects~\cite{Igarashi1999}, drawing lines to depict shapes~\cite{Cole2008}, facial animation~\cite{Miranda2012}, sketching 3D surfaces~\cite{Schmidt2009}, lighting design~\cite{Kerr2009,Karlik2014}, material editing~\cite{Kerr2010,Serrano2016}, icon selection~\cite{Laursen2016}, style similarity~\cite{Garces2014,Garces2016}, or simulation of cloth~\cite{Sigal2015}, to name a few. 
Closer to our approach, Santoni et al.~\cite{Santoni2016} presented a statistical analysis of 3D digital sculpting. The authors analyzed users' behavior as they freely sculpted organic models to discover patterns in the data and trends in the workflows. Regarding the particular case of light fields, Jarabo et al.~\cite{Jarabo:SIG14} presented the first thorough study to evaluate different light field editing interfaces from a user perspective. They identified the local point-and-click operation as the basis for many common edits (such as drawing, increasing brightness or placing additional objects), and performed two different experiments: In the first one, their goal was to compare two interfaces (based on parallax and focus cues, respectively) with respect to their effectiveness, efficiency and subjective preference, by performing several editing tasks of increasing complexity. The conclusions learned guided a second experiment, where users were given the possibility of combining both interface paradigms, and use new tools proposed by the users themselves.

In this paper, we build on this recent work and provide a statistical analysis of subjects' workflows and preferences for each type of editing scenario: editing of surfaces, editing in free space, occlusion handling, and editing of complex geometries. \new{This complements previous work~\cite{Jarabo:SIG14} by focusing on the user's behaviour on the temporal domain, that is, on editing \emph{workflows}, which was previously ignored.}
\\
\new{\paragraph{Signal Processing and Editing} 
Several works on image editing have used signal processing tools for applications such as tone mapping~\cite{Reinhard2002,Durand2002}, contrast enhancement~\cite{Farbman2008,Paris2011}, relighting and recoloring~\cite{Gastal2011}, or material editing~\cite{Boyadzhiev2015}. Many of these works can be coupled with manual intervention to guide the output of such techniques. 
For light field editing, a few works have directly used such processing tools: Most edit propagation techniques for light fields~\cite{Jarabo:SIACG2011,Ao2015,Williem2016} are based on the bilateral filter when defining the propagation energy metric, while some commercial light field software incorporates light field-specific postprocessing filters (e.g., \emph{Lytro Desktop}~\cite{LYTRO} or \emph{Lightfield Iris}~\cite{Iris}). Finally, Garces et al.~\cite{Garces2016intrinsic} expanded the classic Retinex constraint for intrinsic decomposition in light fields. However, in order to act locally or to allow user refinement, these works need precise edit placement; these local operations are the main focus of our work. }

\section{Editing Interfaces and Tools}
\label{section:interfaces_and_tools}
\textbf{Interaction paradigms:} The majority of previous works on light field editing rely on correspondences between the views to specify the desired edits (see for instance~\cite{Seitz2002,Stavrakis:2004:ISP:2383533.2383541,Lo:2010:SCP:1882261.1866173}). This amounts to using parallax as a depth cue to specify the depth at which an edit should be placed. We term this approach \emph{Multiview (M)}. A completely different paradigm based on Isaksen et al.'s light field reparameterization~\cite{Isaksen:2000:DRL:344779.344929} was explored by Davis et al.~\cite{Davis:2012:ULF:2318858.2318864}; depth at which the editing will be performed is specified by a plane of focus, while the rest of the light field is blurred accordingly (wide aperture rendering). This amounts to using blur as a depth cue to specify an edit's position in depth, and we term this approach \emph{Focus (F)}. Thus, both paradigms differ in how the \emph{depth} at which the edit is to be placed is specified. These are the two paradigms originally explored by Jarabo and colleagues{\cite{Jarabo:SIG14}, which we also use for the analysis in this paper.
\\
We now briefly summarize the procedure to perform an edit in the light field with each of our two basic interaction paradigms: Multiview and Focus. Note that we work with \emph{point-based, local edits} (strokes). Our assumption is that a local point-based interaction lies at the basis of the majority of editing processes and tools (from the use of a brush or the eraser to more complex editing tools like selection or global filters). To place a stroke using the \emph{Multiview} interface, the user first draws it in one view, and the epipolar lines of the stroke then become visible. By switching between views and observing the resulting parallax of the stroke, the user then infers its depth, and can move it to the desired position in the light field (Figure~\ref{figure:strokes}, left). Using the \emph{Focus} interface, the user first selects the plane of focus, then draws a stroke on that plane (Figure~\ref{figure:strokes}, right). Notice the main difference in both interaction paradigms: in Multiview mode the stroke is first drawn, and then its position adjusted, while in Focus mode the depth is selected first, and then the stroke is drawn. \\

\begin{figure} [tbp]
   \centering
   \includegraphics[width=\columnwidth]{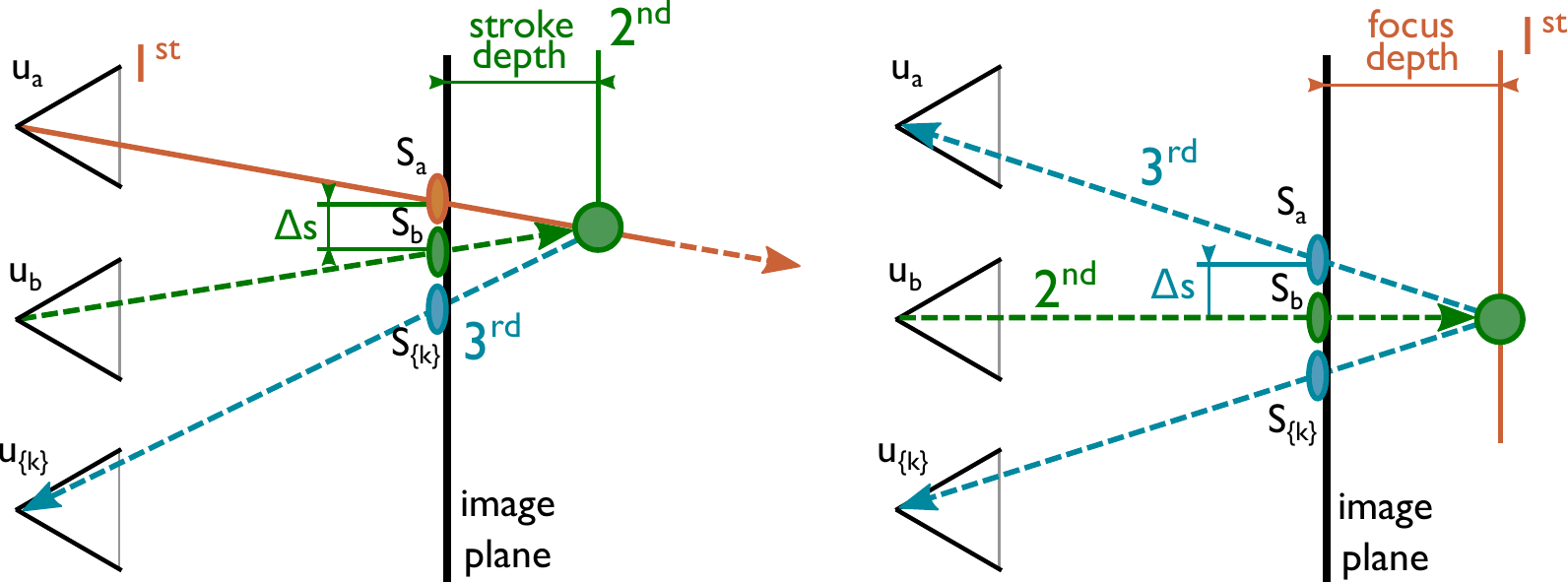}
  \caption{Placing a stroke in the light field using our two interaction paradigms. (a) Multiview: The stroke s$_a$ is first drawn in a view u$_a$ (1$^{st}$), and then its depth adjusted in another view u$_b$ by displacing the stroke along the epipolar line (2$^{nd}$), once this depth is chosen, the stroke is fixed and propagated to other views u$_{\{k\}}$. (b) Focus: First the depth is selected by adjusting the in-focus plane (1$^{st}$), and a stroke s$_b$ is placed in the central view u$_b$ (2$^{nd}$), which is then propagated to the other views u$_{\{k\}}$ (3$^{rd}$). 
  }
\label{figure:strokes}
\end{figure}

\textbf{Interfaces:} The first two interfaces we study are derived from the interaction paradigms described above, which we call \emph{Multiview (M)} and \emph{Focus (F)}. In addition to these, we leverage recent advances in scene reconstruction from light fields (e.g.,~\cite{6247656,Kim:2013:SRH:2461912.2461926}), which allow to infer depth maps, and incorporate that information in two additional interfaces. Figure~\ref{figure:depth_maps} shows example depth maps for two of the light fields used in our tests. The availability of depth information for the input light field may alter the preferred workflow when users perform edits. As such, in the Multiview and Focus interfaces above, we include the option to use depth, yielding the \emph{Multiview with Depth (MD)} and the \emph{Focus with Depth (FD)} interfaces. For both, the strokes drawn will now snap to the nearest surface below them. Therefore, the main difference between these two interfaces lies in the visualization of the light field and the performed edits: While in Multiview with Depth these are visualized by switching the point of view, in Focus with Depth this is done by shifting the in-focus plane. All interfaces share the same screen layout, shown in Figure~\ref{figure:interface}. Details on the implementation of the interfaces can be found in Jarabo et al.~\cite{Jarabo:SIG14}.\\

\begin{figure} [tbp]
   \centering
    \subfloat[Couch]{\includegraphics[width=0.49\columnwidth]{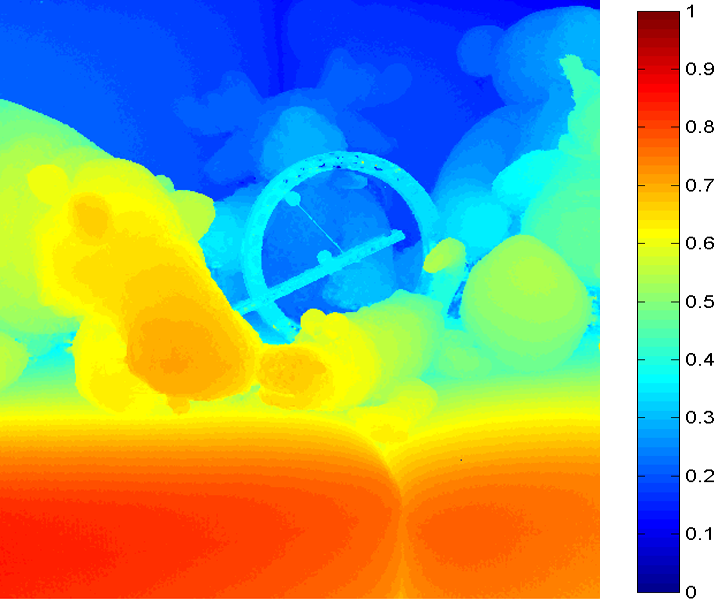}}
	\hspace{\fill}
	\subfloat[Mansion]{\includegraphics[width=0.49\columnwidth]{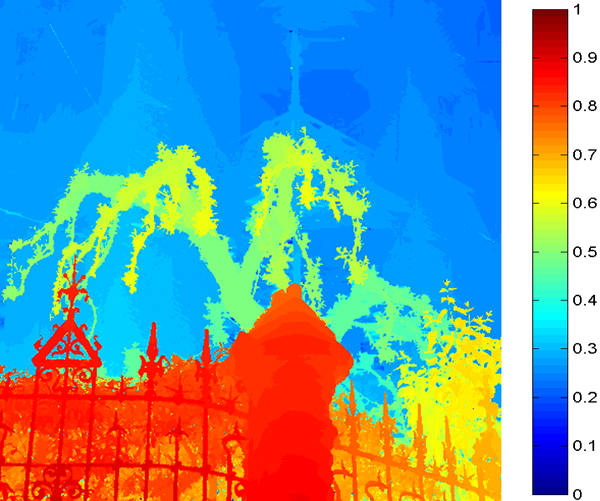}}
  \caption{Depth maps from light fields (central view) in the dataset by Kim et al.~\cite{Kim:2013:SRH:2461912.2461926}, used in the tasks of our second experiments (depth values are encoded as disparity in pixels).}
\label{figure:depth_maps}
\end{figure}

\begin{figure} [tbp]
   \centering
    \subfloat[Mutiview]{\includegraphics[width=0.49\columnwidth]{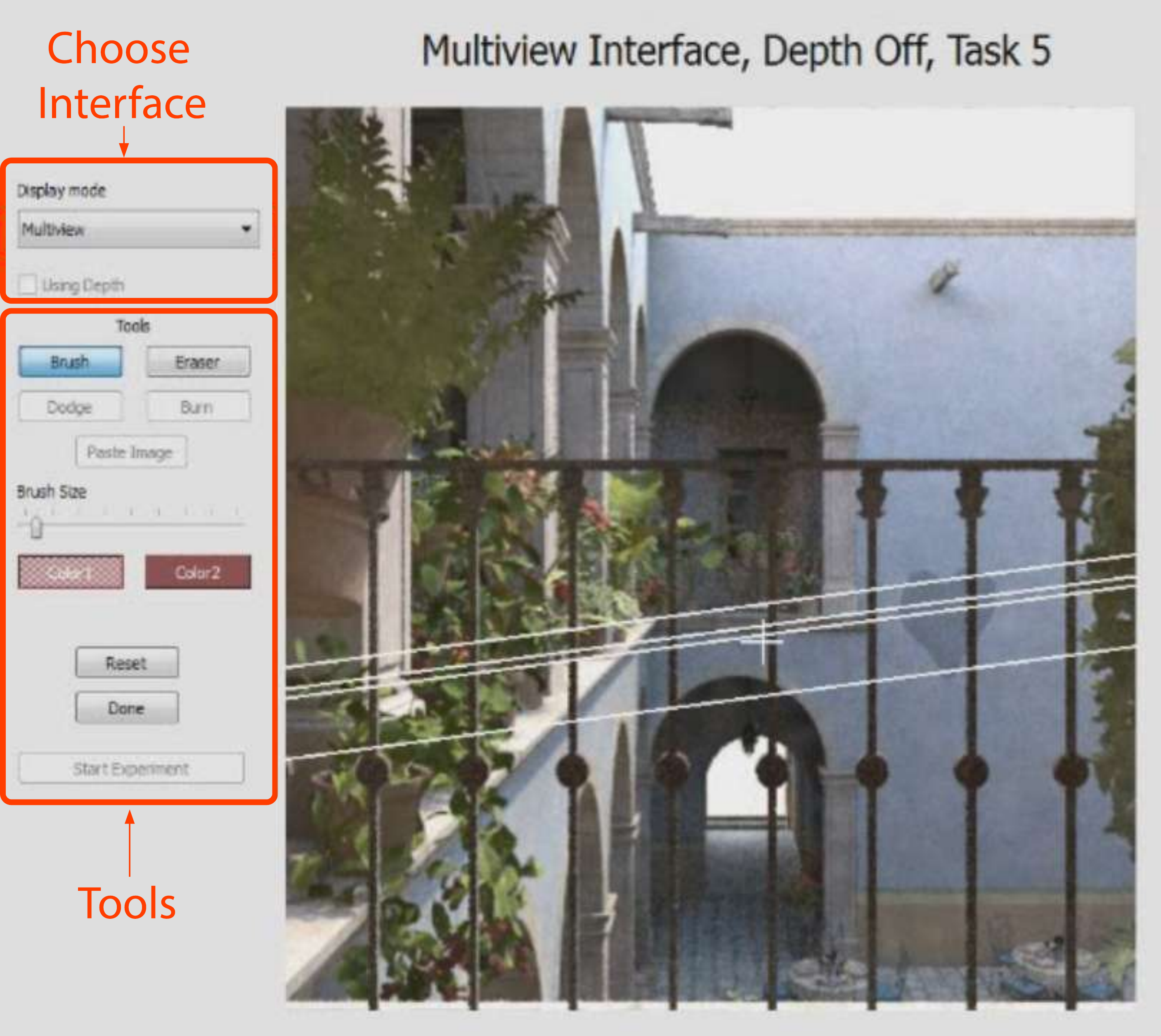}\label{figure:interface_multiview}}
	\hspace{\fill}
	\subfloat[Focus]{\includegraphics[width=0.49\columnwidth]{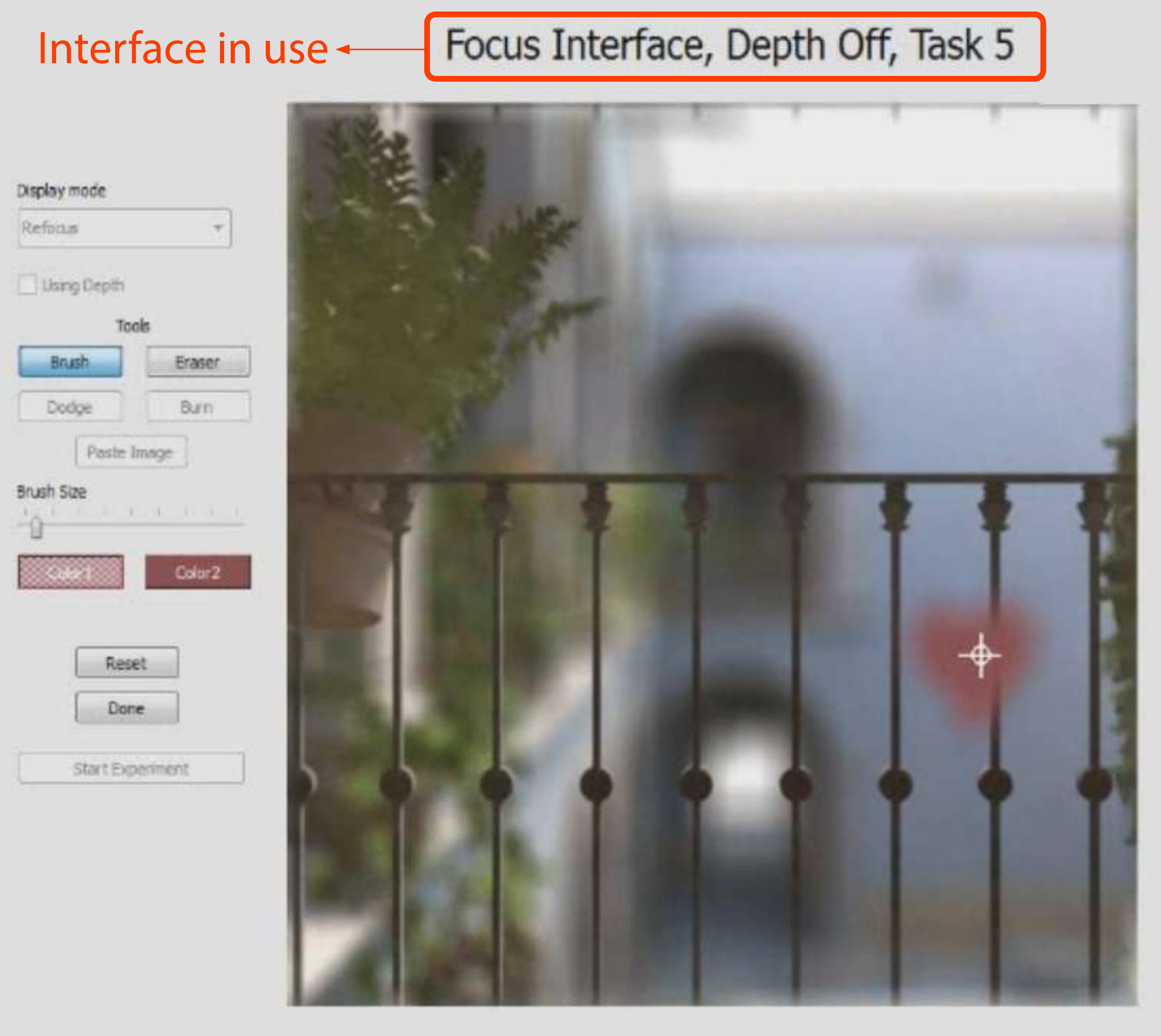}\label{figure:interface_focus}}
    \caption{User interfaces used in our tests. In the Multiview paradigm, epipolar lines (white) mark the trajectory a currently active stroke will follow when moved in depth. In the Focus paradigm, the in-focus plane marks the depth at which strokes will be placed.}
\label{figure:interface}
\end{figure}


\textbf{Editing tools:} Our interfaces incorporate the following set of basic tools: draw, erase, and pasting of pre-loaded billboard objects. To navigate the light field, they include: change view (only available in the Multiview interfaces) and set depth (only available in the interfaces without depth). In the Focus interfaces, depth is navigated using the mouse wheel. 
We further include a \emph{depth selection} tool, which allows the user to pick a certain depth (mouse click), and specify a depth range around it using a slider. Edits can then only be performed inside the specified depth range; this is similar to a common mask tool, but working in depth space. Similar in spirit, a \emph{color selection} tool is also included. With it, the user selects a color and a threshold, creating a corresponding selection mask. Finally, we incorporate a \emph{visual aid} tool, which highlights the selected areas of the light fields by overlaying a semi-transparent checkerboard pattern on all other, non-selected areas. 

For a better understanding of the interfaces and tools available we refer the reader to the videos of editing sessions which can be found online\footnote{{\url{http://giga.cps.unizar.es/~ajarabo/pubs/lfeiSIG14/#downloads}}}.

\section{Experimental Design} 
\label{section:experimental_design}
We briefly describe the two experiments carried out, whose results are thoroughly analyzed in the following sections. In every case, both objective and subjective data was collected. Objective data comprised timing data, tools used, or measures of error in depth when available. Subjective data included ratings and rankings by users collected via questionnaires and free form comments. More detailed information can be found in Jarabo et al.~\cite{Jarabo:SIG14}.  

\subsection{Synthetic Scenarios (Experiment 1)}
\label{subsection:experiment1}
\textbf{Stimuli:} We used two synthetic light fields for the different tasks: a complex architectural scene (\emph{San Miguel}), and a still-life scene (\emph{Vase}) (Figures~\ref{figure:experiment}a and~\ref{figure:experiment}b). 
Since they are synthetic scenes, the actual ground truth geometry and depth are known.
\\

\textbf{Tasks:} The first experiment contains five tasks [S1..S5], performed in fixed order. They include a target image, and subjects were given a detailed explanation of the desired edit. 
Tasks were chosen so that subjects would perform a variety of edits, including editing planar (S1) and curved surfaces (S2), emphasizing highlights (S3), placing objects in free space (S4), and dealing with occlusions (S5). Two examples are depicted in Figure~\ref{figure:experiment}. 
\\

\textbf{Procedure:} Twenty paid subjects (6 female, 14 male) took part in the experiment. \new{This is consistent with the number of subjects typically used in other similar statistical user studies, which usually range between two and 30~\cite{Santoni2016,Schmidt2009,Karlik2014,Kerr2010}.} 
Each subject used each of the four interfaces for all tasks, in random order to avoid possible learning effects. For each interface, tasks were performed in fixed order, S1 to S5. Time per task was limited to five minutes. 

\begin{figure} [tbp]
   \centering
    \subfloat[Draw on wall]{\includegraphics[width=0.3\columnwidth]{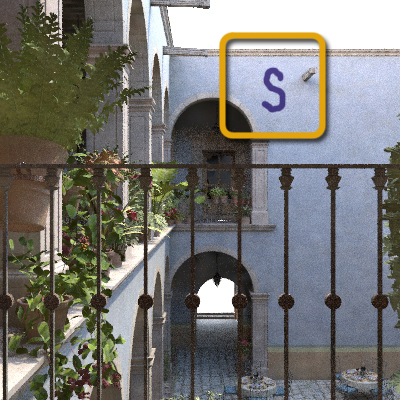}\label{figure:S1}}
	\hspace{\fill}
	\subfloat[Edit highlights]{\includegraphics[width=0.3\columnwidth]{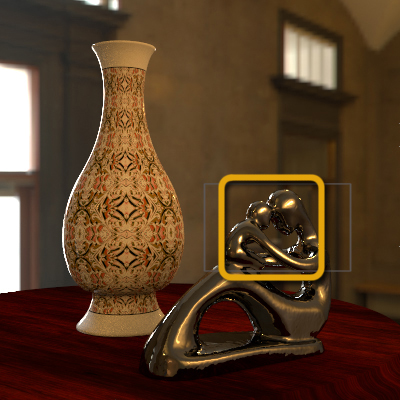}\label{figure:S3}}
	\hspace{\fill}
	\subfloat[Add ivy to the wall]{\includegraphics[width=0.3\columnwidth]{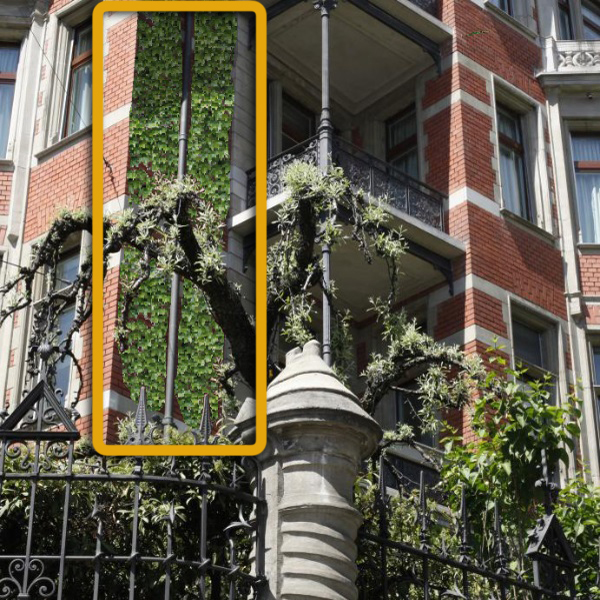}\label{figure:R5}}
  \caption{Target image and description for three representative editing tasks (from left to right, S1 and S3 from Experiment 1, and R5 from Experiment 2). For images of the remaining editing tasks, please refer to~\cite{Jarabo:SIG14}.}
\label{figure:experiment}
\end{figure}

\subsection{Real Scenarios (Experiment 2)}
\label{subsection:experiment2}
In Experiment 2, we used a hybrid interface combining all the different possibilities offered by the original four interfaces in Experiment 1, and used real, captured light fields. 
A key difference between the synthetic light fields used in Experiment 1 and real light fields is the absence of ground-truth depth information; instead, we are limited by the depth recovered by the camera, which is not error-free. 
\\

\textbf{Stimuli:} We used eight real light fields (an example is shown in Figure~\ref{figure:experiment}c) captured by a Lytro$^{\text{TM}}$ light field camera, a camera gantry, or a Raytrix$^{\text{TM}}$  camera. They were chosen to cover a variety of scenes, depth, and reflectance complexities. 
\\

\textbf{Tasks:} This experiment contains ten tasks [R1..R10], in which a target image and a detailed explanation of the editing task to perform were given to subjects. The tasks were: colorization on slanted (R1) and curved surfaces (R2), cloning an object (R3), altering the appearance of a material (R4), adding texture (R5), importing a billboard object (R6), changing luminances (R7), altering color and placing a logo (R8), tweaking small details (R9), and harmonizing the color of the scene (R10). Again, tasks were chosen to cover a wide variety of edits. Figure~\ref{figure:experiment}c shows the target image given to subjects for one of the tasks in Experiment 2.
\\

\textbf{Procedure:} Ten different subjects (4 female, 6 male) took part in the experiment. 
Each subject was presented the ten tasks in random order to avoid possible learning effects. To carry out the tasks, they could choose freely between a Multiview and a Focus paradigm, and between using or not depth information. Time per task was limited to ten minutes. 


\section{Interface Suitability Based on the Task to Perform}
\label{section:task_analysis}

This section discusses the suitability of each interface for the different editing tasks. We focus on Experiment 1, where we can easily compare how tasks are performed with each interface. Analysis of the data for time to completion, error in depth (measured as the L1 norm averaged across views), and ratings and rankings on interface preference provided by users yield three distinct clusters, roughly corresponding to three task categories: editing of surfaces (planar or curved), editing in free space, and occlusion handling. Data analysis was performed using repeated measures ANOVA for error, timings, and ratings, and Kruskal-Wallis for rankings~\cite{Cunningham2011}. 
\\

\textbf{Editing of surfaces:} Tasks S1 to S3 allow us to draw insights into surface editing. In these tasks, error in depth for interfaces with depth (MD and FD) is zero, since strokes snap to the surface below them (Figure~\ref{figure:depth_error}). Consequently, realizing the task with these interfaces took less time (Figure~\ref{figure:time_to_completion}). For interfaces without depth, M yielded a higher error than F, showing that users found it more difficult to locate an edit in depth with M. This is also reflected in the timings, in which M is significantly slower. 
\\

\begin{figure*} [tbp]
   \centering
     \subfloat[]{\includegraphics[width=\columnwidth]{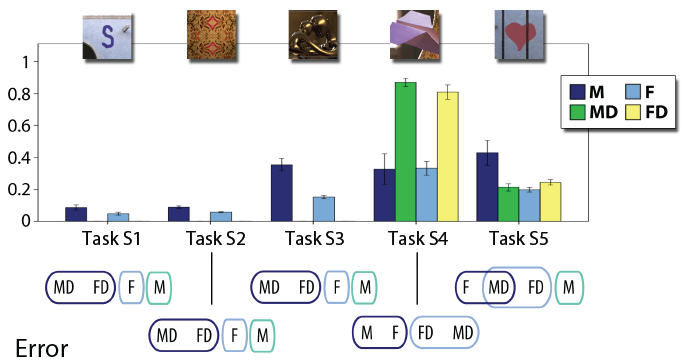}\label{figure:depth_error}}
	\hspace{\fill}
	\subfloat[]{\includegraphics[width=\columnwidth]{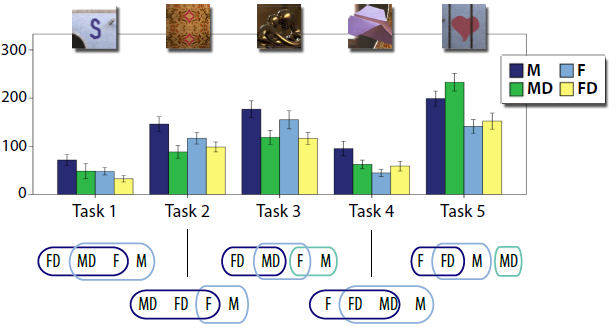}\label{figure:time_to_completion}}
  \caption{\textbf{Left:} Mean error in depth for the five directed tasks in Experiment 1, and results of the pairwise comparisons between interfaces. \textbf{Right:} Same for time to completion. Source:~\cite{Jarabo:SIG14}.}
\label{figure:results_experiment1}
\end{figure*}


\textbf{Editing in free space:} Task S4 deals with positioning in free space. In this case, interfaces without depth yield lower error than those with depth, and the difference between interaction paradigms is not significant (Figure~\ref{figure:depth_error}). Even though errors are high, times to completion are very low (Figure~\ref{figure:time_to_completion}). In interfaces with depth, this can be due to the fact that users realized that they would not be able to correctly place it in depth (recall that the edit will snap to the surface right below) and gave up. This hypothesis seems supported by the low ratings these two interfaces received in this task.
F is the interface of choice among the four tested for editing in free space, because it provides clear depth feedback and results in shorter editing times.
\\

\textbf{Occlusion handling:} Task S5 implies handling occlusions. For this, F yields both the lowest error and time to completion (Figure~\ref{figure:results_experiment1}), and its superiority is confirmed by ratings. 
\\

In summary, this experiment has shown the large influence of the task to perform in the interface of choice, with, essentially, MD and FD being ahead for on-surface editing, and F the interface of choice for free-space editing and occlusion handling.

\section{Workflow Analysis}
\label{section:workflow_analysis}

Data collection from the experiments described in Section~\ref{section:experimental_design} yields an immense quantity of both subjective and objective information, which provides insights into a variety of different aspects of light field editing: usage of different tools, preferred interfaces, variability of preferences with the task to perform, workflows, etc. While previous work has focused on how suitable different interfaces and tools are, and whether an interface can allow for satisfactory editing of light fields with current depth reconstruction methods~\cite{Jarabo:SIG14}, here we focus on subjects' workflows, that is, we look for underlying patterns in subjects' actions and their preferences for different generic tasks. \new{As far as we know, this is an unexplored area of light field editing in which our intuitions on how users approach the editing tasks still need to be formally demonstrated.}

\new{Systematically analyzing the temporal activity of subjects while editing is a challenging task. We choose to model this data as categorical sequences, i.e., as a sequence of discrete possible states. States can either correspond to tools (Sec.~\ref{subsection:tool_sequence_analysis}) or to interfaces (Sec.~\ref{subsection:interface_sequence_analysis}), depending on what we want to analyze. There are multiple techniques in the literature to analyze categorical sequences, including correspondence analysis of states~\cite{Deville1983}, or simply applying sequence alignment algorithms and computing pairwise distances~\cite{Abbott2000}. In this work, we use several state sequence analysis techniques, including Markov modeling. We briefly explain here the fundamentals of these techniques, and then move on to describe the findings derived from them, both for tools and interfaces.}

\new{For the state sequence analysis---performed using the R library TraMineR~\cite{TraMineR}---, we first present visualizations of both individual and transversal statistics. The former refer to individual sequences, the latter present aggregated information, such as the state distribution plot, which shows, for each time instant, the aggregated distribution of states. More importantly, we then move on to compute \emph{representative} sequences. One of the main concerns in state sequence analysis is to analyze the (dis)similarity between sequences and try to summarize subjects' behavior in a few possible sequences. A possible representative sequence is the medoid of the set, i.e., the sequence for which the sum of distances of the sequences in the set is minimized~\cite{TraMineR}. However, just one sequence does not adequately model the high variability present in cases like editing; we therefore compute a set of representatives, as it is explained below.
Finally, to look into transitions we resort to Markov modeling~\cite{Berchtold2002}. We extract first-order Markov chains to analyze the transitions between states and determine which are the most common local workflows. Then, to further analyze more complex workflows, we search for higher-order hidden Markov sequences, which allows us to evaluate whether this local behavior holds over longer sequences.}



\subsection{Tool Sequence Analysis (Experiment 1)}
\label{subsection:tool_sequence_analysis}

We assume that the tools used in the editing process correspond to different states: \emph{draw}, \emph{erase}, \emph{change view}, and \emph{set depth}. We include an \emph{idle} state to represent that no tool is being used, and a \emph{finished} state. We assume that idle periods shorter than one second correspond to switching from one tool to another; therefore, we eliminate them from the tool sequence. In this analysis, editing workflows are grouped by interface since it was preset and users could not switch among the four available ones.
\\

\textbf{Tool sequence visualization:} \emph{Sequence index plots} render state sequences and allow us to observe the succession of states and, through the length of each color segment, the time spent in each state. Figure~\ref{figure:task3_sequences} represents the sequence of tools used by the 20 participants to complete Task S3 (edit specular highlights). 
While using M and MD, users constantly change view to check the appearance of the edits. This indicates that they are not approaching light field editing the way they approach conventional image editing; rather, they are very aware of the high dimensional nature of the scene. The use of change view is even more extensive when depth is not present, since they use it to place edits at a correct depth. This typically increases editing times: the time to complete the task tends to be longer when no depth is present. This holds for both F and M: a large amount of the time is spent on correct depth placement of edits (using the set depth tool to move the stroke along epipolar lines in M, or to set the desired plane in focus in F), which seems a strong advocate for the use of depth on light field interfaces. Interfaces featuring depth also register an increased time spent drawing (without an increase in erasing, which actually decreases, as we will show later in Figure~\ref{figure:meanTime_per_tool}a), further indicative of the fact that users can concentrate more on the task at hand and its accurate completion.
\\

\begin{figure}[tbp]
   \centering
   \includegraphics[width=1\columnwidth]{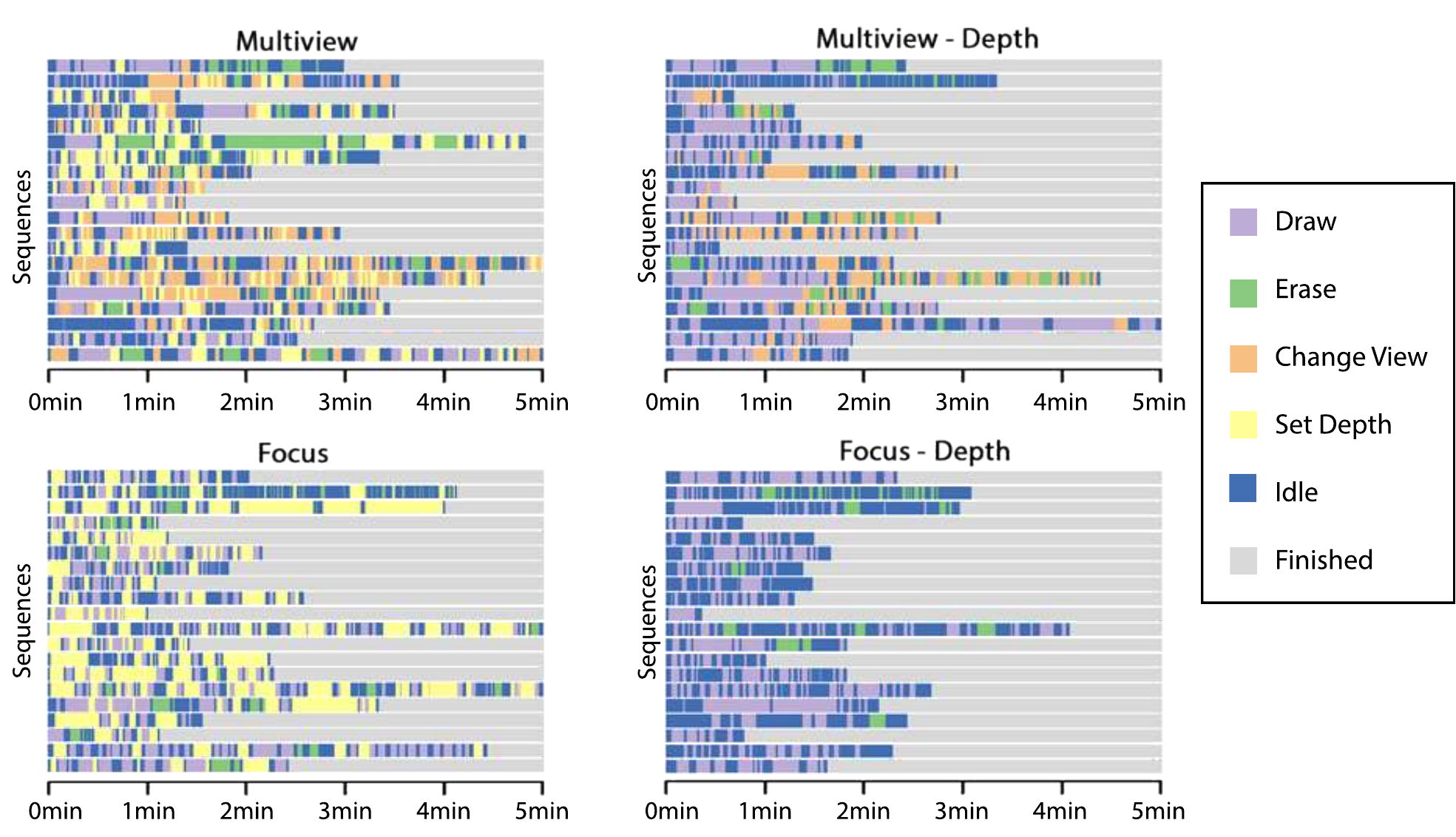}
  \caption{Sequence index plot for Task S3 in Experiment 1. The horizontal bars show the sequence of tools the 20 participants used to complete the task with each one of the four interfaces, each tool being represented with a different color.}
\label{figure:task3_sequences}
\end{figure}

\textbf{Mean time using each tool:} Figure~\ref{figure:meanTime_per_tool} shows the average amount of time, not necessarily consecutive, spent using each tool for Tasks S3 (edit specular highlights) and S4 (place object in free space). Note that, as  explained before, the change view tool is only available in M, while the set depth tool can only be used when depth is off. 
Comparing Tasks S3 and S4 we also notice how different tools are involved in different tasks: while Task S3 requires drawing and erasing, in Task S4 the user only had to paste a new object, resulting in minimal drawing and erasing. 
At the same time, the comparison between such different tasks allows us to see patterns that are common despite the nature of the task.
We observe that the time spent setting the depth is consistently longer in F. This is because modifying the depth is used first to choose the plane we want to edit, and then to check the result, which is costly for the user, resulting in longer editing times. In M, the depth is only used to adjust the stroke position. View changes are used to check results in both M and MD, and in the former also for stroke adjustment in depth. This generates a more frequent use of the change view tool for M compared to MD.
\\

\begin{figure*} [tbp]
   \centering
    \subfloat[Task S3. Edit specular highlights.]{\includegraphics[width=\columnwidth]{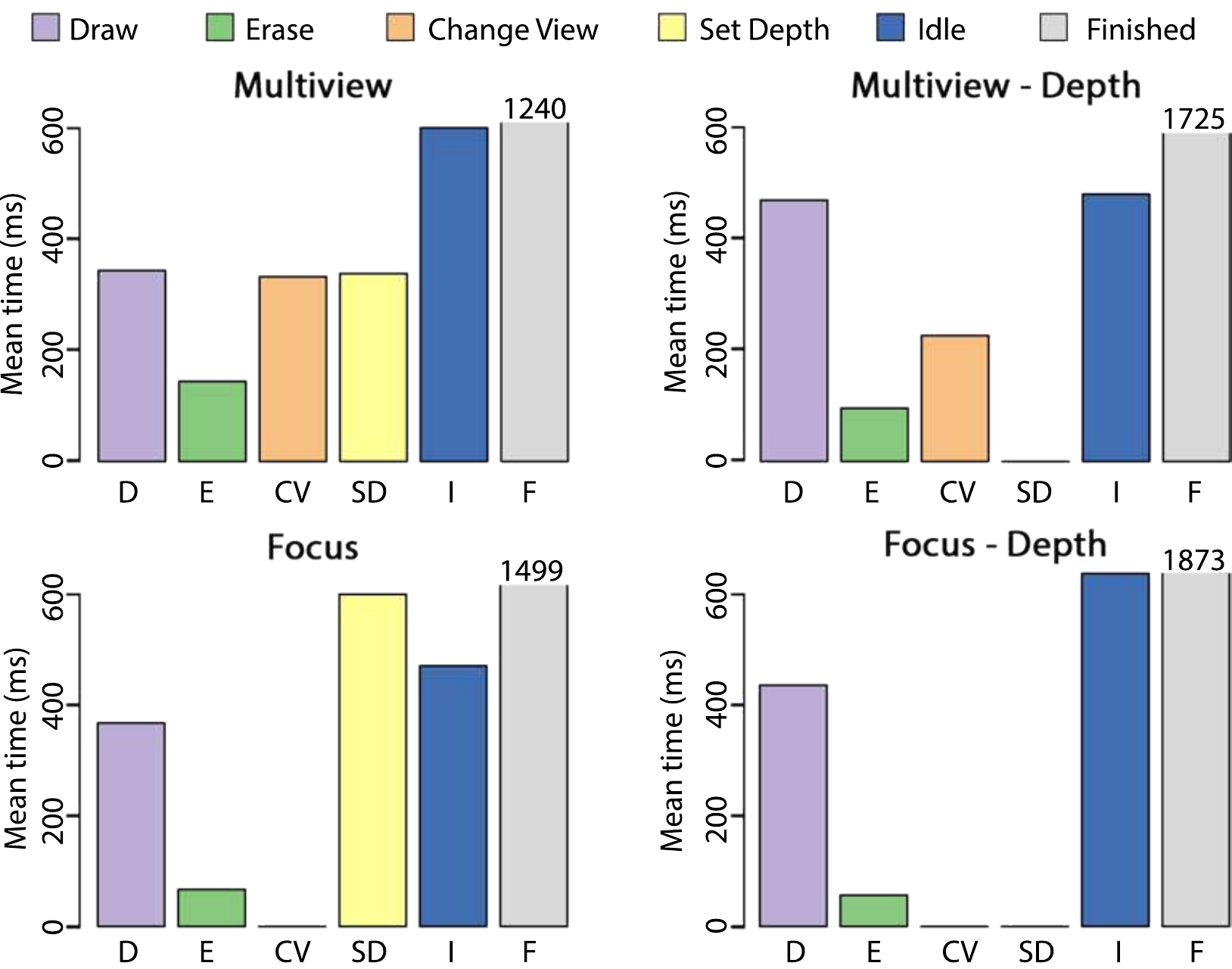}\label{figure:meantime_TASK3}}
	\hspace{\fill}
	\subfloat[Task S4. Add airplane in free space]{\includegraphics[width=\columnwidth]{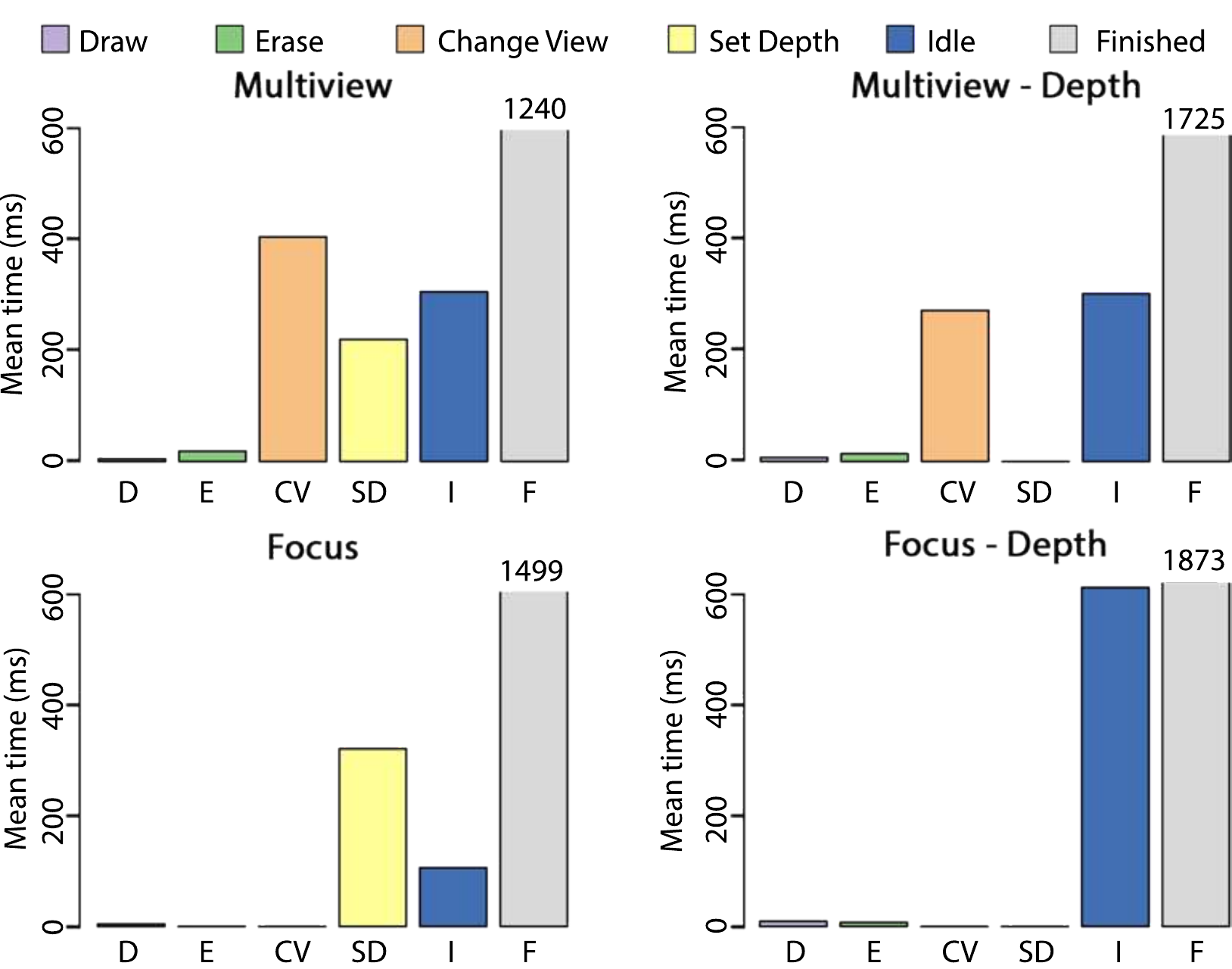}\label{figure:meantime_TASK4}}
  \caption{Mean time using each tool for Tasks S3 and S4 in Experiment 1. Results are grouped by interface; in reading order: Multiview, Multiview with Depth, Focus, and Focus with Depth.}
\label{figure:meanTime_per_tool}
\end{figure*}

\textbf{Tool Usage Distribution:} A \emph{state distribution plot} displays the general pattern of a whole set of state sequences. Unlike sequence index plots, they do not render individual sequences, but provide an aggregated view of the frequency of each tool for each time interval. Figure~\ref{figure:merged_statedistribution} shows the tool usage distribution for all the tasks in Experiment 1. This represents how the use of different tools is distributed along time when varied editing tasks are performed. For example, in FD, at the beginning about 40\% of the users are drawing while the rest are idle. After a while, a percentage of users erase strokes, and all the participants start to gradually finish the editing task.

Changing the view is more frequent when depth is not activated, since it is needed to adjust the position of strokes in different views, as opposed to MD, where it is only used to check results. Users spend more time in idle state in FD, probably because they are moving the cursor on the screen observing how the focus changes and deciding where to draw.

The main conclusion we extract from these plots is that there seems to be no clear or preferred order of states in any of the interfaces. The existence of a preferred pattern would mean, for instance, that at the beginning a majority draws, then a majority changes view to check results, then a majority erases mistakes, etc. However, the plots show that throughout time the distribution of the non-finished people among the available states remains constant, e.g, in MD, shortly after the beginning, there is a 1/3-1/3-1/3 distribution among draw-change view-idle that approximately remains true throughout time. However, we will see in the next paragraph that underlying patterns of actions can be found if shifting in time is considered when looking for them.
\\

\begin{figure}[tbp]
   \centering
   \includegraphics[width=\columnwidth]{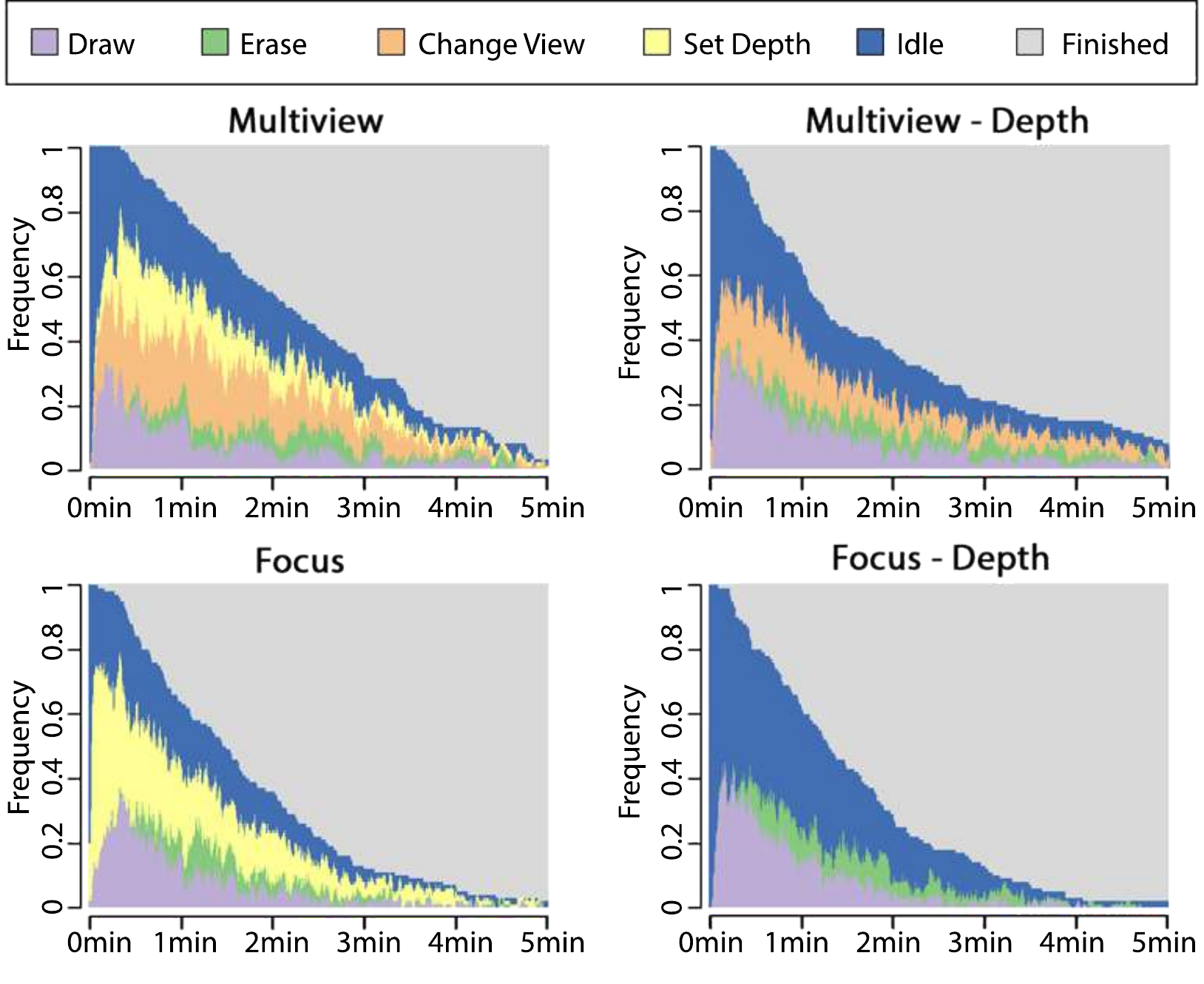}
  \caption{Tool usage distribution for all tasks in Experiment 1 grouped by interface. Note that graphs do do not represent individual sequences, they show the frequency of use of each tool at each time interval.}
\label{figure:merged_statedistribution}
\end{figure}

\begin{figure*} [tbp]
   \centering
\includegraphics[width=\textwidth]{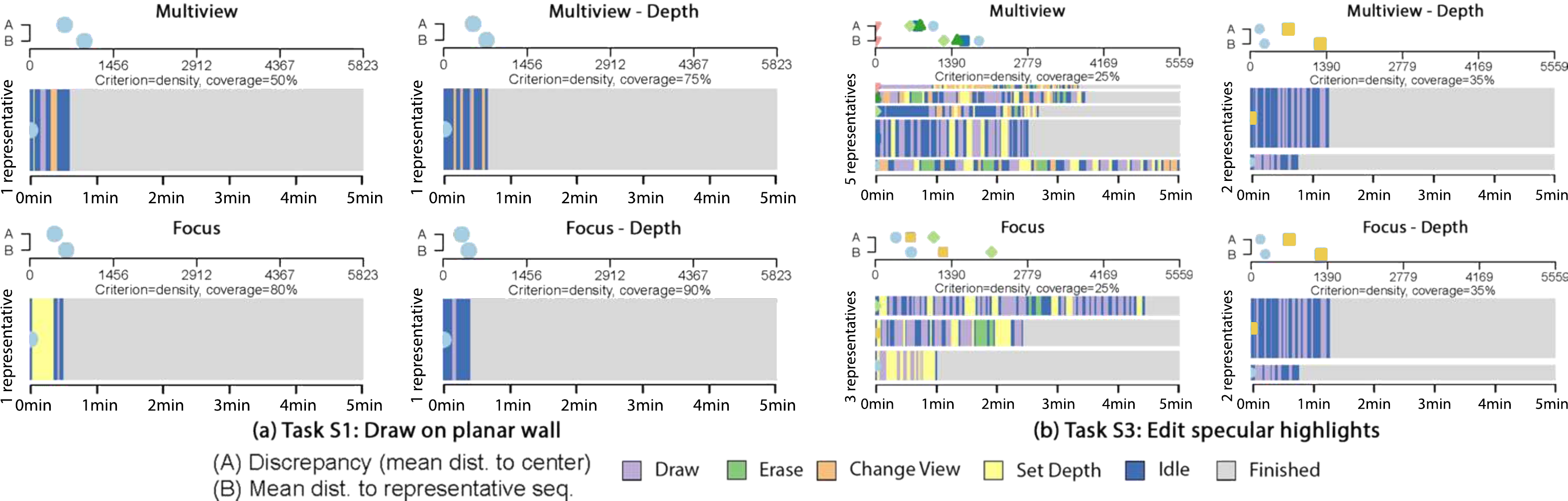} 
  \caption{Representative tool sequences for Tasks S1 and S3 in Experiment 1 grouped by interface with a minimum coverage of 25\% and Optimal Matching distance used to compute dissimilarities. Representative sequences are plotted bottom-up according to their representativeness score and the bar width is proportional to the number of sequences assigned to them. Above each plot, two values are displayed for each representative: the discrepancy within the subset of sequences assigned to that representative (A), and the mean distance from the sequences to the representative (B). The scale ranges from zero to the maximum theoretical distance.}
\label{figure:representative_sequences}
\end{figure*}

\textbf{Representative Tool Sequences:} Once we allow time shifting in the search for patterns, we find that a large number of editing sequences can be summarized with a representative set. Figure~\ref{figure:representative_sequences} shows the set of representative sequences for Tasks S1 and S3 grouped by interface. The sets cover at least 25\% for the sequences in each one of the groups, meaning that the obtained sequences are representative of at least a quarter of the total number of sequences. In order to select the representatives, we compute the pairwise dissimilarities among sequences by calculating the Optimal Matching (OM) distance~\cite{OptimalMatching}. OM gives us the minimum cost of transforming one sequence into another allowing two transforming operations: the substitution of one element by another, and the insertion/deletion of an element, which generates a one-position shift of all the elements to its right. We have computed this metric with an insertion/deletion cost of one and a substitution cost matrix based on observed transition rates.

Comparing the representative tool sequences for a given task with the four interfaces, we can see clear differences. The sequence followed in FD is typically composed of idle and drawing intervals, while in MD, changing view is also used.
It is interesting to note that, if depth is off, variability among users is much higher (Figure~\ref{figure:representative_sequences}, right); this is especially the case in M. This may indicate increased hesitation and experimenting: the path to completion is less clear, and users need to experiment more and end up following different paths.

Comparing two different tasks we notice that some of them can be edited by different users following many different sequences, while other (simpler) tasks are edited very similarly by all users: in Task S4, two to five representative sequences are required to obtain 25\% coverage (Figure~\ref{figure:representative_sequences}, right); instead, in Task S1 only one representative covers 50\% or more of the sequences (Figure~\ref{figure:representative_sequences}, left).
\\

\begin{figure} [tbp]
   \centering
    \subfloat[Multiview]{\includegraphics[width=0.5\columnwidth]{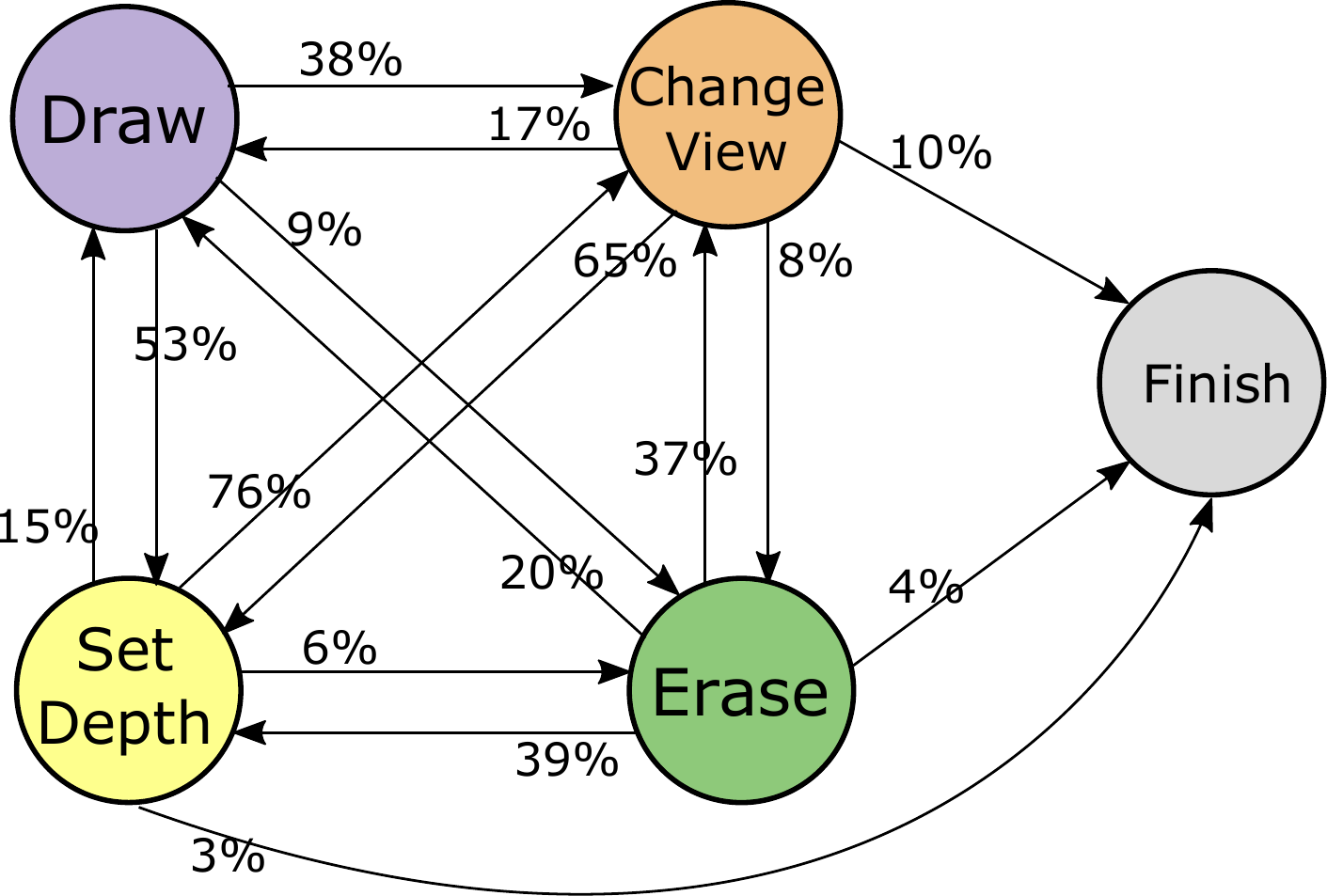}\label{figure:transitions_multiview}}
    \hfill
    \subfloat[Multiview with Depth]{\includegraphics[width=0.5\columnwidth]{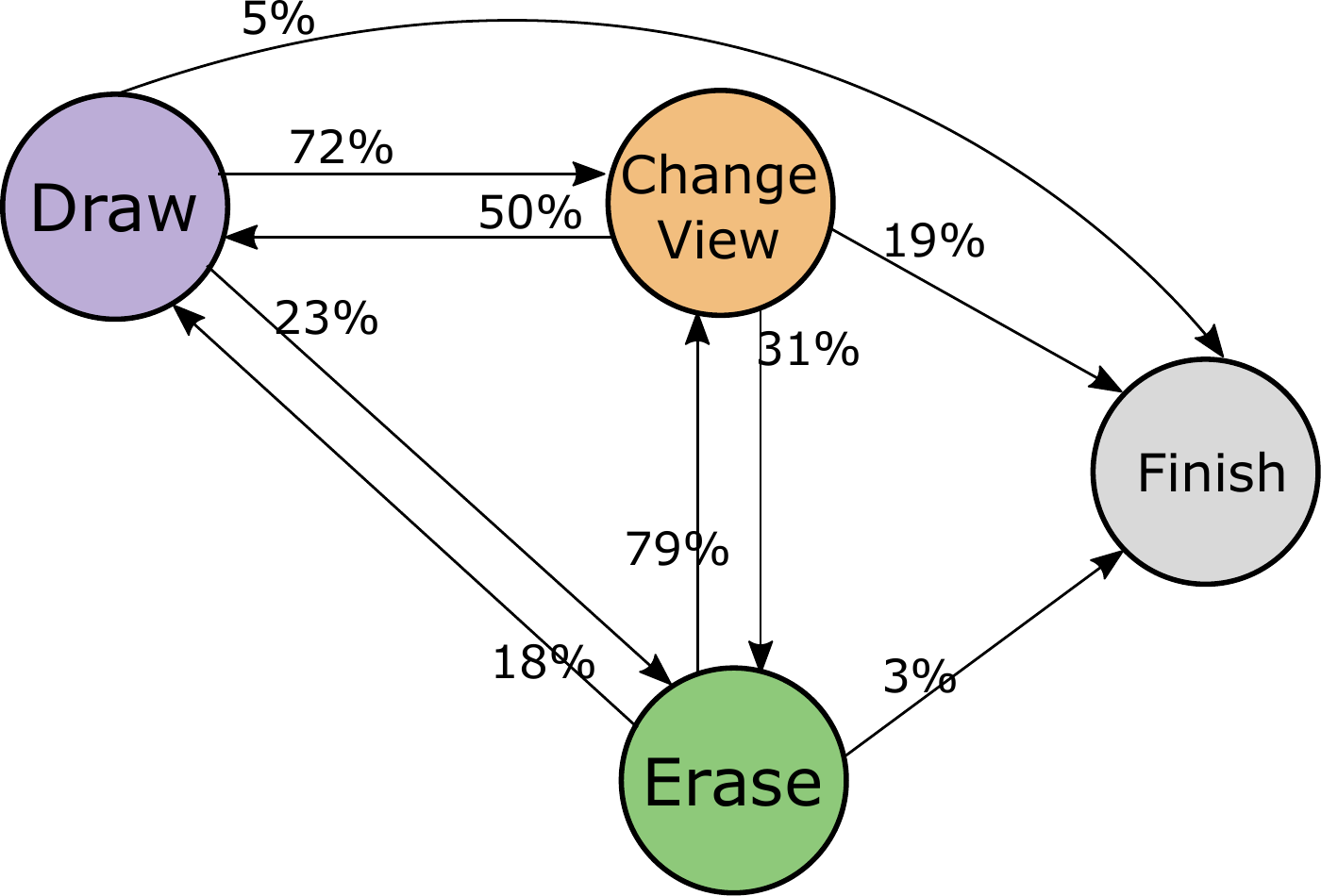}\label{figure:transitions_multiview_depth}}\\
    \subfloat[Focus]{\includegraphics[width=0.5\columnwidth]{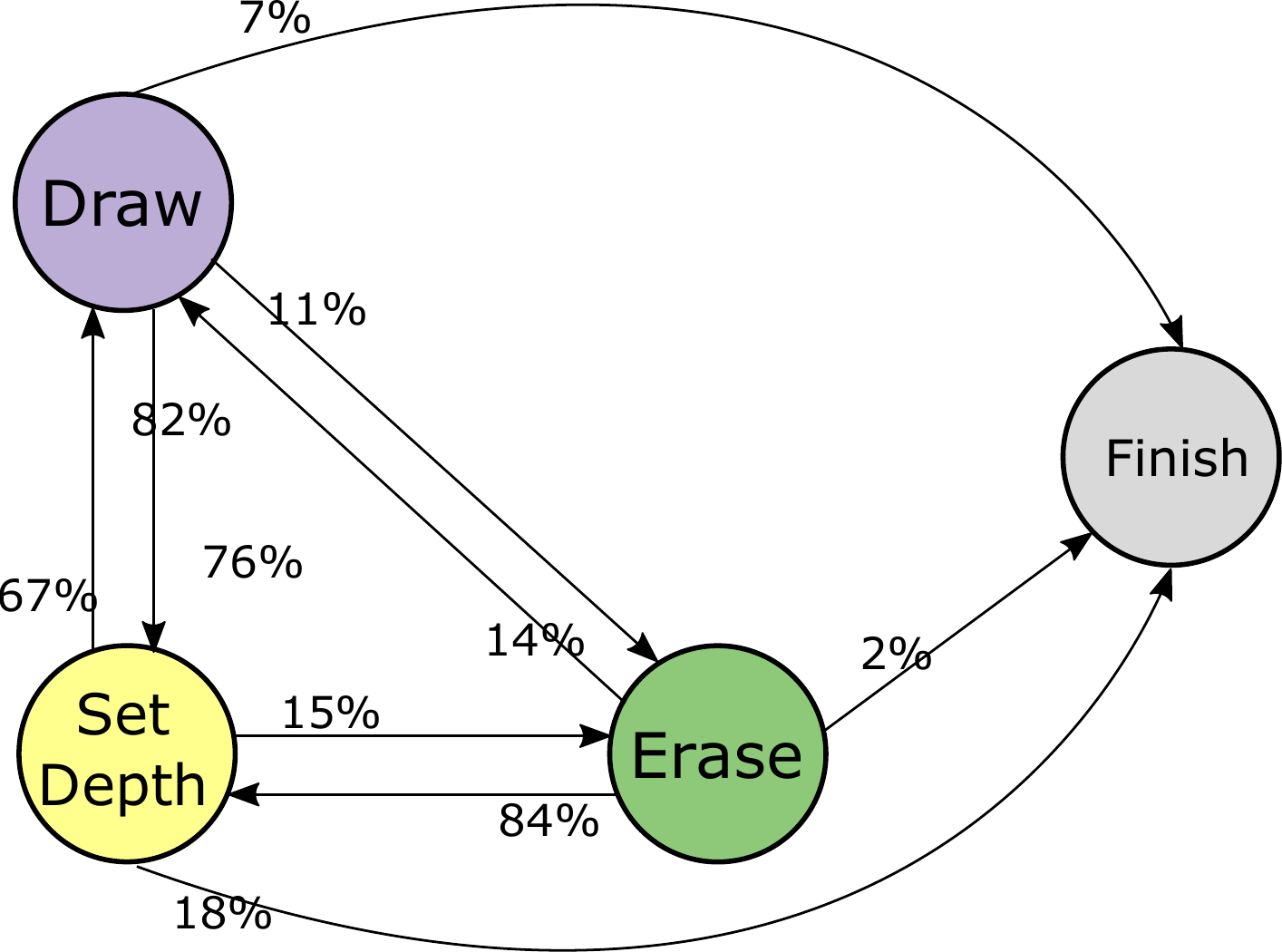}\label{figure:transitions_focus}}
    \hfill
    \subfloat[Focus with Depth]{\includegraphics[width=0.5\columnwidth]{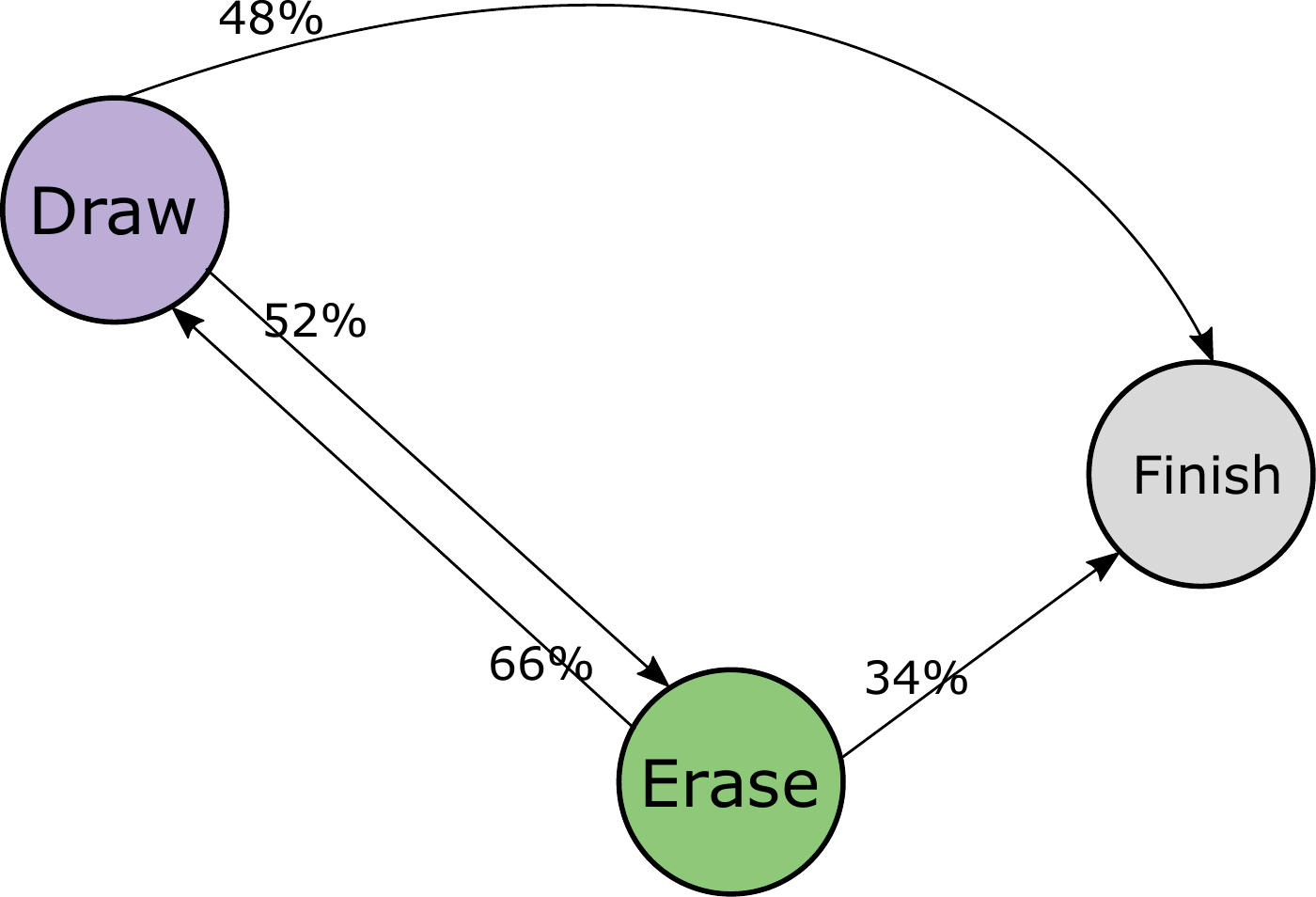}\label{figure:transitions_focus_depth}}
  \caption{Effective tool transition probabilities for all the tasks in Experiment 1. Self-loop transitions have been removed and the idle state is not considered. Note that set depth only appears when depth is not activated. With depth on, edits are automatically placed in the surface below the cursor.}
\label{figure:tool_transitions}
\end{figure}

\textbf{Effective Tool Transitions:} \emph{Transition rates} between each couple of states give us the probability of switching from one state to another. In image editing, users always spend long periods of time on the same task (e.g., drawing), and normally are idle for a while when they change from one tool to another. To avoid that behavior from biasing our results, we consider only \emph{effective transitions}, that is, we eliminate self-loop transitions and remove the idle state. Figure~\ref{figure:tool_transitions} presents the probability of the effective transitions for all the tasks in Experiment 1, grouped by interface.

In M (Figure~\ref{figure:transitions_multiview}), the editing workflow consists of constantly changing the view to observe the appearance of strokes and setting the depth to move them along the epipolar lines, with occasional drawing and erasing to perform small corrections. We also notice that users never finish editing directly after drawing, because they have to check how the new strokes look in different views first. In MD (Figure~\ref{figure:transitions_multiview_depth}), strokes automatically snap to the surface below the cursor, so adjusting depth is not necessary. Participants change the view only to check their edits, and draw and erase to fix the errors they detect. Again, most transitions to finish come from change view, indicating that users use it to check the result.

With F (Figure~\ref{figure:transitions_focus}), setting the depth is extensively used in combination with drawing and erasing. The large number of transitions from draw and erase to set depth and from set depth to finish shows that users not only use it to set the desired plane in focus but also to check the results. In this case, the view cannot be changed. Finally, when using FD (Figure~\ref{figure:transitions_focus_depth}) neither change view nor set depth are available. The user must simply draw and erase until (s)he obtains the desired effect, and can check the results by moving the cursor on the screen to set the region below the mouse in focus.

In order to further analyze these workflows, we extract the hidden Markov chains (MC) up to order five (chains with six states), which complement the first order MC shown in Figure~\ref{figure:tool_transitions}. This analysis shows that, for M, the most common workflow in all tasks consists of looping between setting the edits' depth and navigating through the light field views to check whether the edit is correctly placed. Besides, users generally navigate before finishing the task. On the other hand, the limited navigation capabilities of F simplifies the workflows: in general, users loop between setting the depth and drawing. Similar to M, the users usually navigate through the light field by using the only available degree of freedom, which in this case is depth focus. In the cases where depth is available (MD and FD), the workflows are significantly simplified, and users focus mostly on surface editing (drawing, and then erasing for refining). However, in MD, users still make heavy use of navigation to check the positioning of the edits and their correct propagation across the light field views. Again, in most occasions users check the correctness of the edits by navigating before finishing the task. 

Overall, we make two main observations: First, the workflow is a constant iteration of acting upon the scene (drawing/erasing) and checking the results, which is common among artists working with other media too~\cite{Santoni2016}. Our interfaces thus seem to offer a valid approach for the usual artist workflow to be applied to light fields. Second, our interfaces allow users to leverage the high dimensionality of light fields during editing, instead of finding it cumbersome to navigate.

\subsection{Interface Sequence Analysis (Experiment 2)}
\label{subsection:interface_sequence_analysis}

\begin{figure*} [tbp]
   \centering
    \subfloat{\includegraphics[width=1.3\columnwidth]{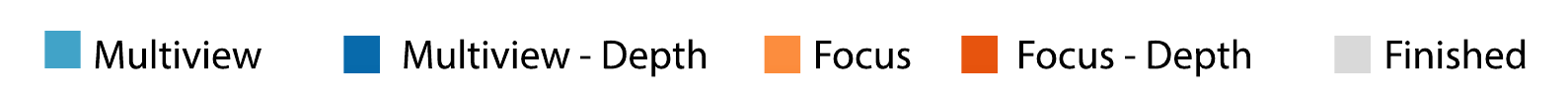}}
	\hspace{\fill}\\
	\renewcommand{\thesubfigure}{a}
	\subfloat[Task R1 (\emph{editing surfaces})]{\includegraphics[width=0.49\columnwidth]{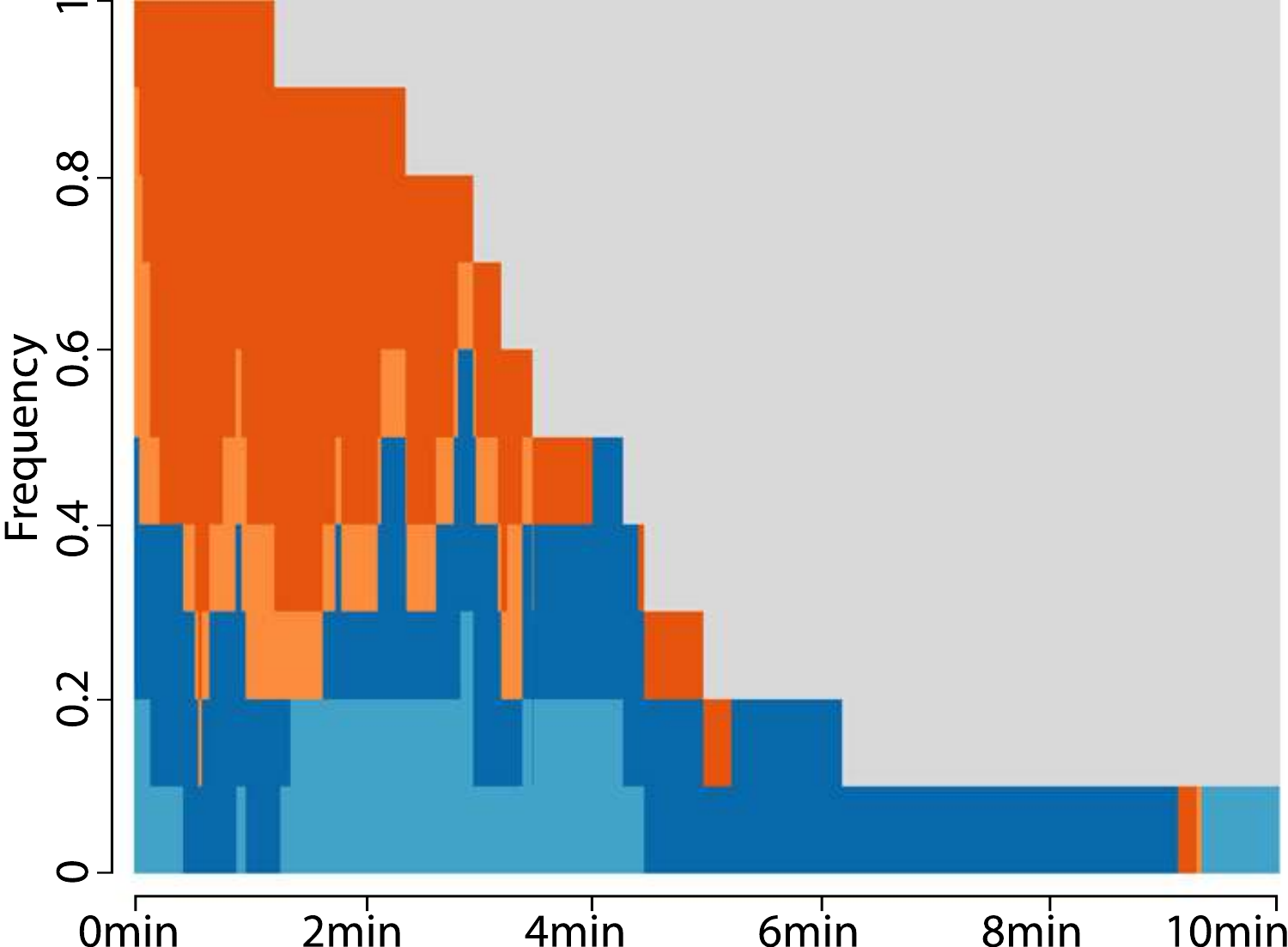}\label{figure:distribution_R1}}
	\hspace{\fill}
	\renewcommand{\thesubfigure}{b}
    \subfloat[Task R3 (\emph{editing in free space})]{\includegraphics[width=0.49\columnwidth]{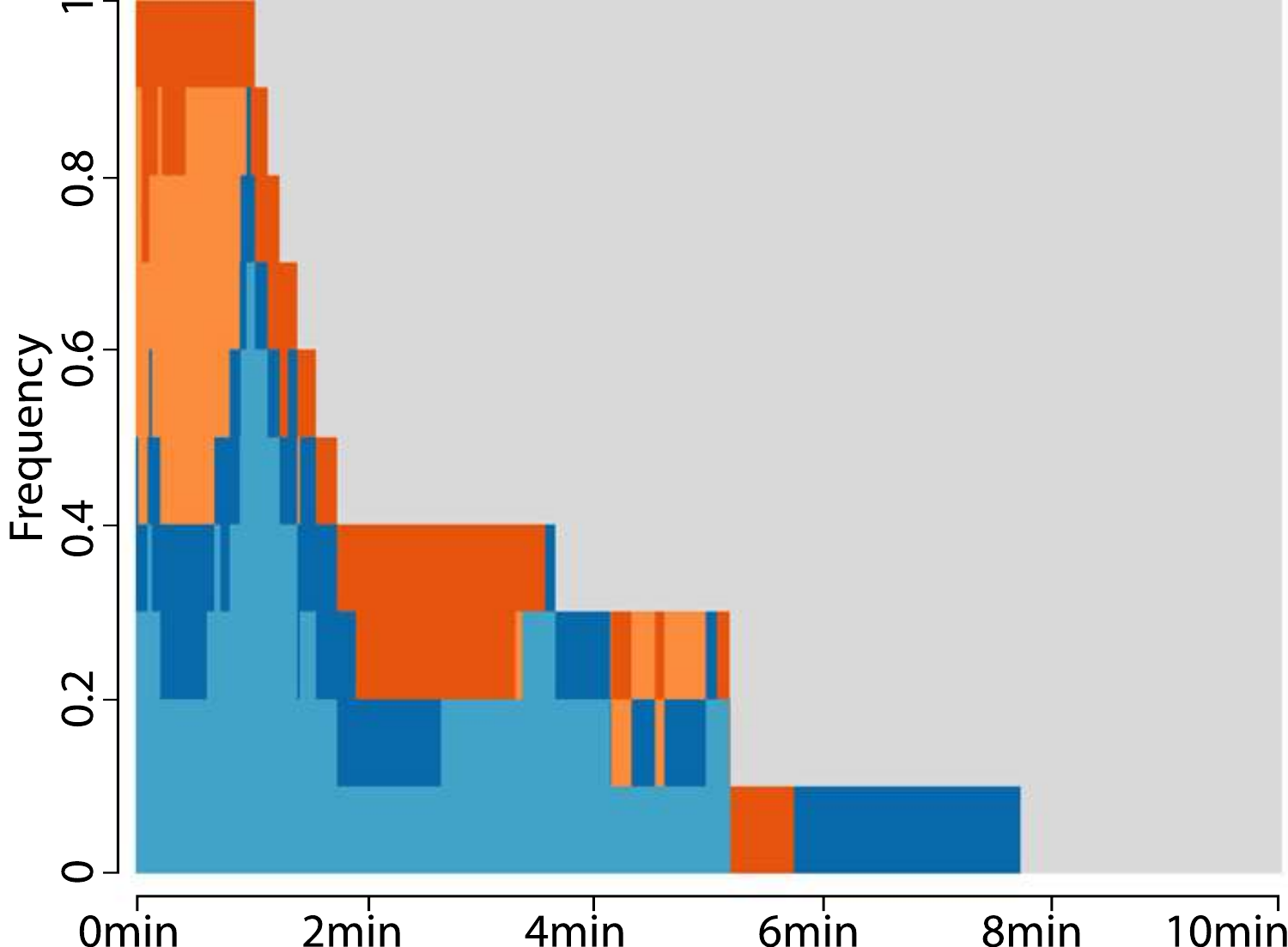}\label{figure:distribution_R3}}
	\hspace{\fill}
	\renewcommand{\thesubfigure}{c}
    \subfloat[Task R5 (\emph{handling occlusions})]{\includegraphics[width=0.49\columnwidth]{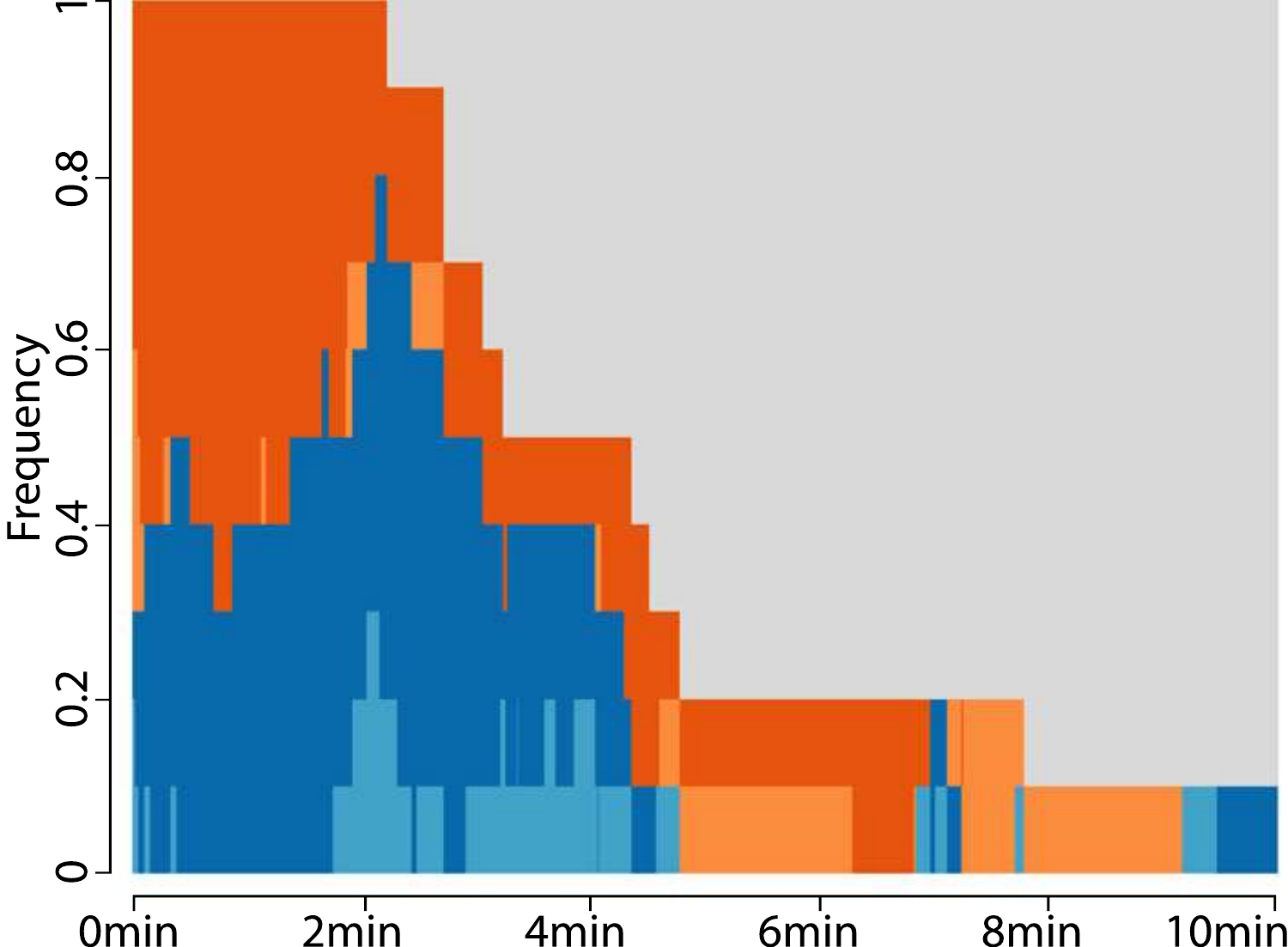}\label{figure:distribution_R5}}
	\hspace{\fill}
	\renewcommand{\thesubfigure}{d}
    \subfloat[Task R10 (\emph{complex geometries})]{\includegraphics[width=0.49\columnwidth]{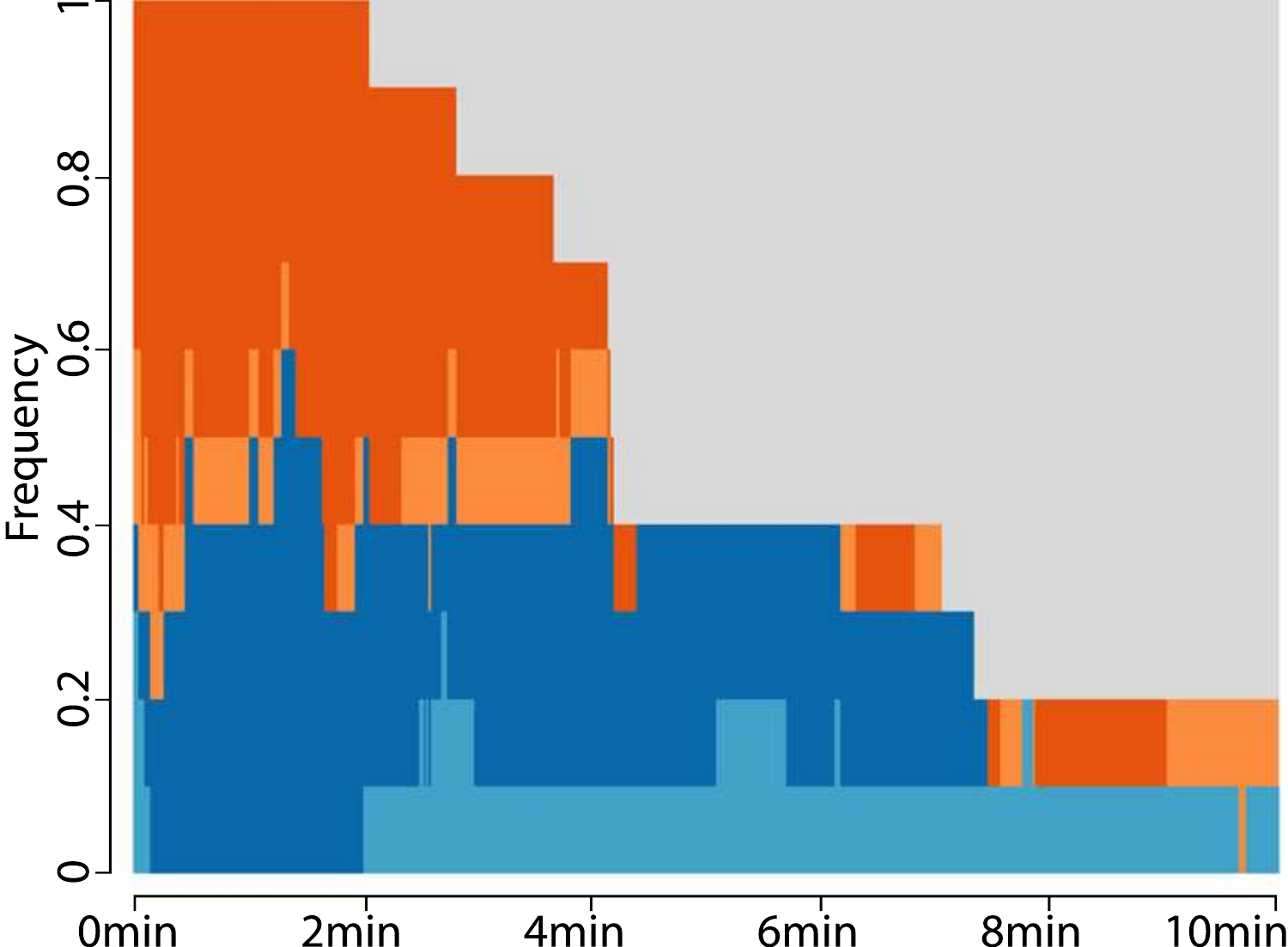}\label{figure:distribution_R10}}
  \caption{Interface usage distribution for Tasks R1, R3, R5, and R10, one pertaining to each of the four categories or use case scenarios (editing surfaces, editing in free space, handling occlusions, and editing complex geometries). Note that graphs do do not represent individual sequences, they show the frequency of use of each interface at each time interval.}
\label{figure:interface_distribution}
\end{figure*}

\begin{figure*} [tbp]
   \centering
   \includegraphics[width=\textwidth]{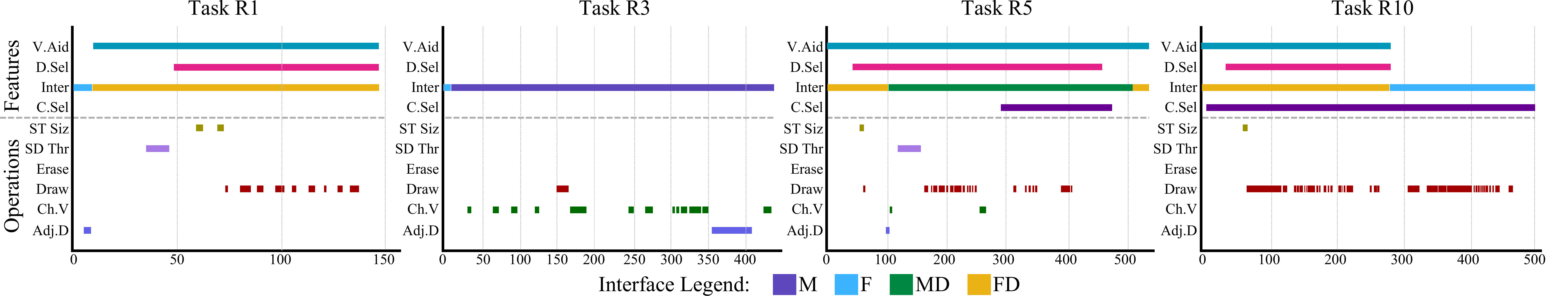}
  \caption{Sample workflows for Tasks R1, R3, R5 and R10, one pertaining to each of the four editing categories (editing surfaces, editing in free space, handling occlusions, and editing complex geometries). They indicate the tools and interfaces used by the subject along time, shown in the abscissa (in seconds). The six bottom marks in the y-axis correspond to different operations; from bottom to top: set depth (\emph{Adj.D}), change view (\emph{Ch.V}), draw, erase, set depth threshold (\emph{SD Thr}), and set tool size (\emph{ST Siz}). The top four marks indicate whether that feature was activated or not at each time instant, and correspond, from bottom to top, to color selection tool (\emph{C.Sel}), interface (\emph{Inter}), depth selection tool (\emph{D.Sel}) and visual aid tool (\emph{V.Aid}). Note that \emph{Inter} features different colors specifying the interface being used. These workflows for all subjects and tasks can be found at \protect\url{http://giga.cps.unizar.es/~ajarabo/pubs/lfeiSIG14/}.}
\label{figure:workflows}
\end{figure*}

In this experiment, users can freely choose which interaction paradigm (Focus or Multiview) to use during the editing process, as well as whether they want to use depth information or not. Further, they can change among these interfaces as often as they want during the editing process. Additionally, in this experiment there are two new tools available to users, namely the color and the depth selection tools (please see Section~\ref{section:interfaces_and_tools} for detailed explanations).
Therefore, we now include the use of these interfaces considering five states: \emph{Multiview}, \emph{Multiview with Depth}, \emph{Focus}, \emph{Focus with Depth}, and \emph{finished}. As mentioned above, our focus here is on workflows, so we look at times of use of the different interfaces and tools as an indicator of preferences. We also look into transitions between states, and additionally collect data on subjective preferences.

We cluster the tasks in categories (editing of---planar or curved---surfaces, editing in free space, and occlusion handling), with an additional category for editing objects of intricate geometry, for which subjects follow a different workflow than when editing simpler planar or curved surfaces. We choose four tasks as an example of each one of the four categories: R1, R3, R5, and R10, respectively.
\\

\textbf{Interface Usage Distribution and Sample Workflows:} 
\old{Tasks R1, R2, R7, R9, and R9 involve editing planar and curved surfaces. In all of these tasks, the use of depth information is largely favored, which matches our results in Experiment 1 and can be seen in the representative example depicted in Figure~\ref{figure:distribution_R1}. There is no clear preference between interaction paradigms (F or M), although there is a slight trend towards F; in the debriefing interviews, subjects reported that Focus offered a very strong and easy-to-interpret cue for visualization of the active area. Regarding the tools used, Tasks R2 and R9 favor the use of the color selection tool, possibly because the areas requiring editing are small, similar in color, and without a distinct depth with respect to their surrounding areas. The rest favor the use of the depth selection tool. A sample editing workflow for Task R1 is shown in Figure~\ref{figure:workflows}.}

\old{Task R3 requires editing in free space. In this case, again, results are consistent with Experiment 1: depth information is scarcely used (Figure~\ref{figure:distribution_R3}). We observe here a trend towards Multiview, as shown also in the workflow for Task R3 included in Figure~\ref{figure:workflows}. This is possibly due to the absence of high frequency information in the area to edit, which causes the blur of the Focus interface to provide little or no depth cues.}

\old{In this experiment, Tasks R5 and R6 require dealing with occlusions. Here, the introduction of the depth and color selection tools causes a change with respect to results obtained in Experiment 1. While in the first experiment there was a large amount of erasing to deal with the occlusions, the introduction of the depth selection tool, largely used in both R5 and R6, reduces the need to erase to a minimum (see Figure~\ref{figure:workflows}, Task R5, for a sample editing workflow in that task). Surprisingly, there is little difference between the use of MD and FD, revealing that as long as depth information and related tools are present, the interaction paradigm is less relevant for these tasks. The color selection tool is fairly used in Task R5 to avoid the pipe, which is hard to disambiguate from the rest of the wall in the depth dimension.}

\begin{figure*} [tbp]
   \centering
    \subfloat[Task R1 (\emph{editing surfaces})]{\includegraphics[width=0.5\columnwidth]{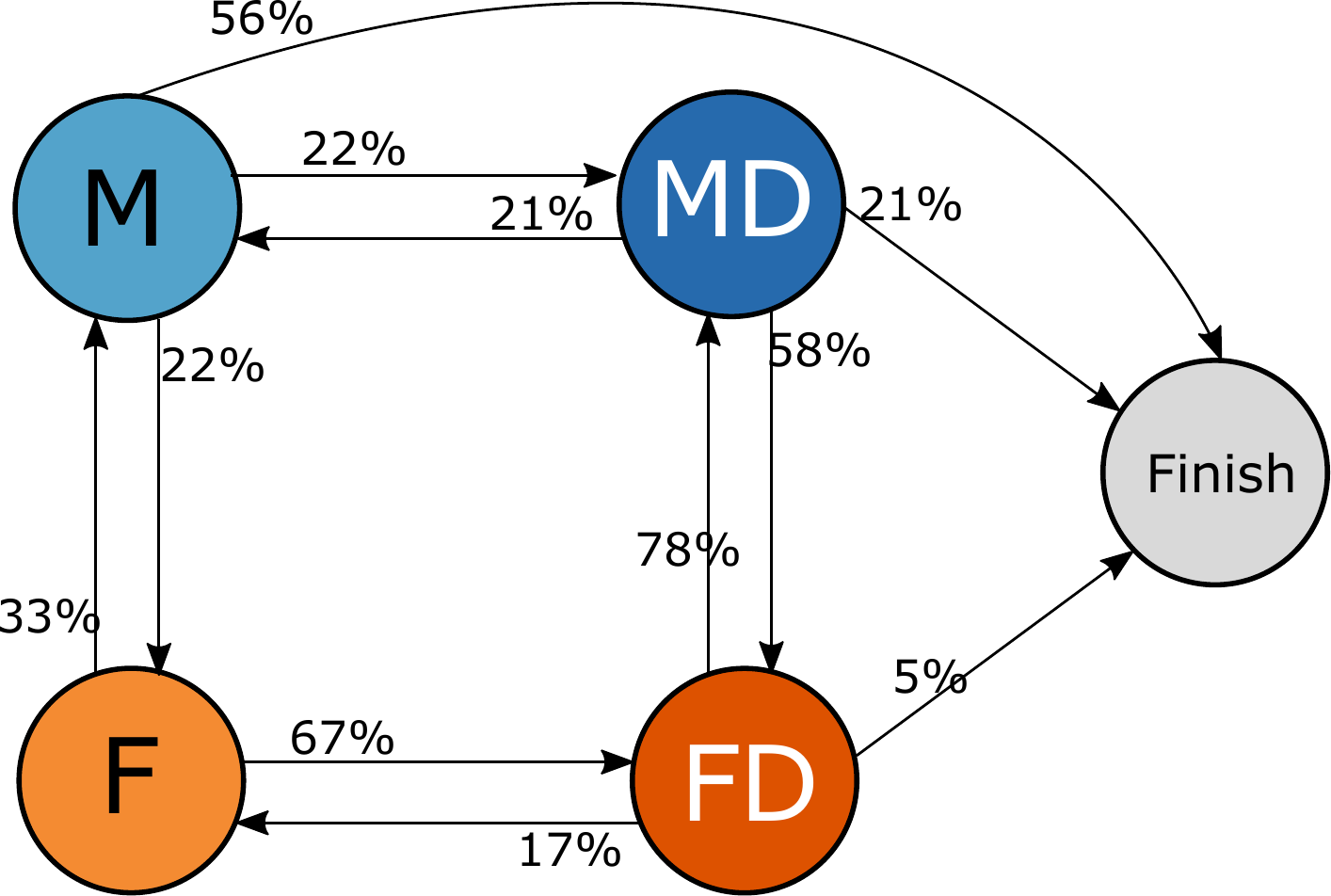}\label{figure:transitions_r1}}
    \hfill
    \subfloat[Task R3 (\emph{editing in free space})]{\includegraphics[width=0.5\columnwidth]{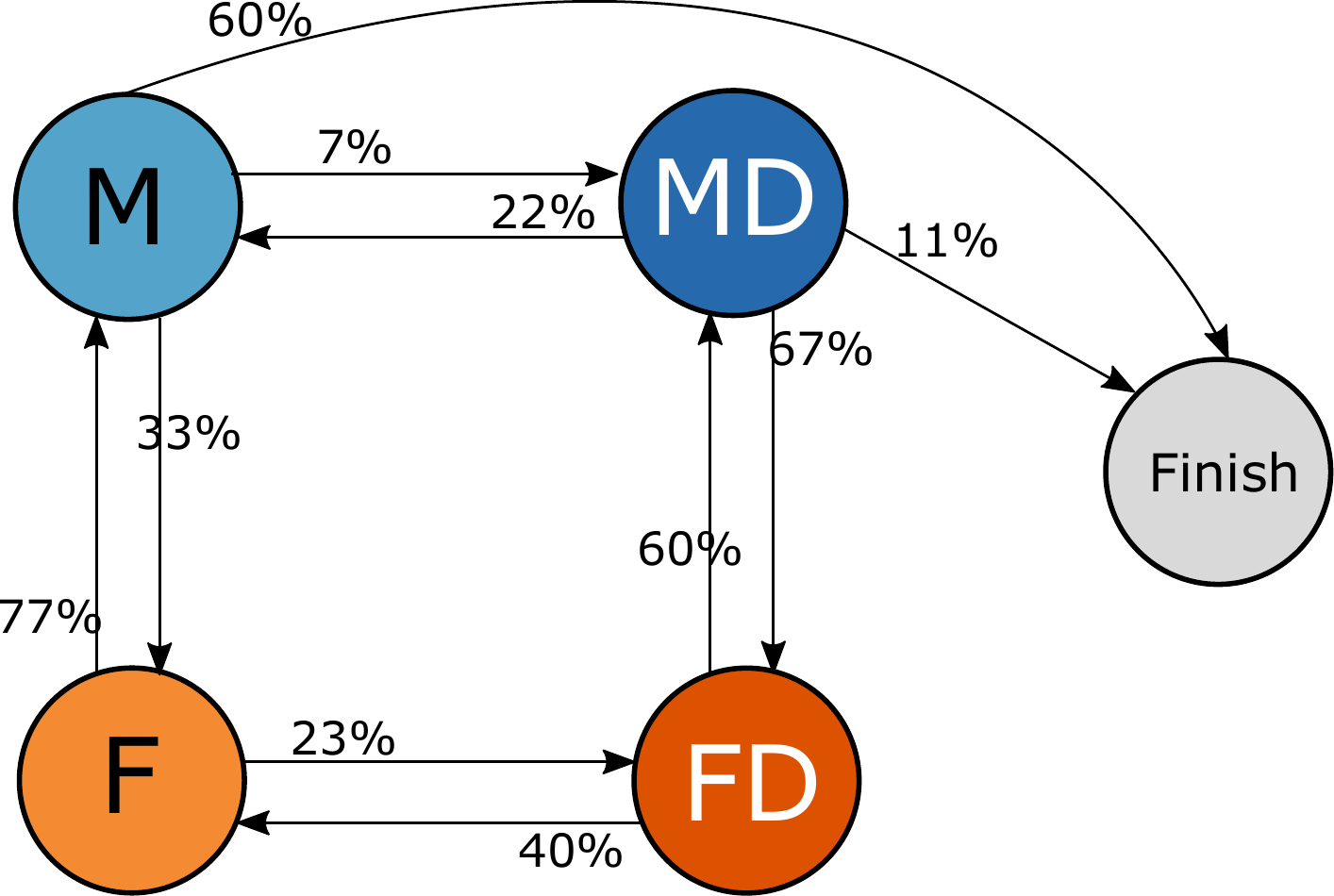}\label{figure:transitions_r3}}
    \hfill
    \subfloat[Task R5 (\emph{handling occlusions})]{\includegraphics[width=0.5\columnwidth]{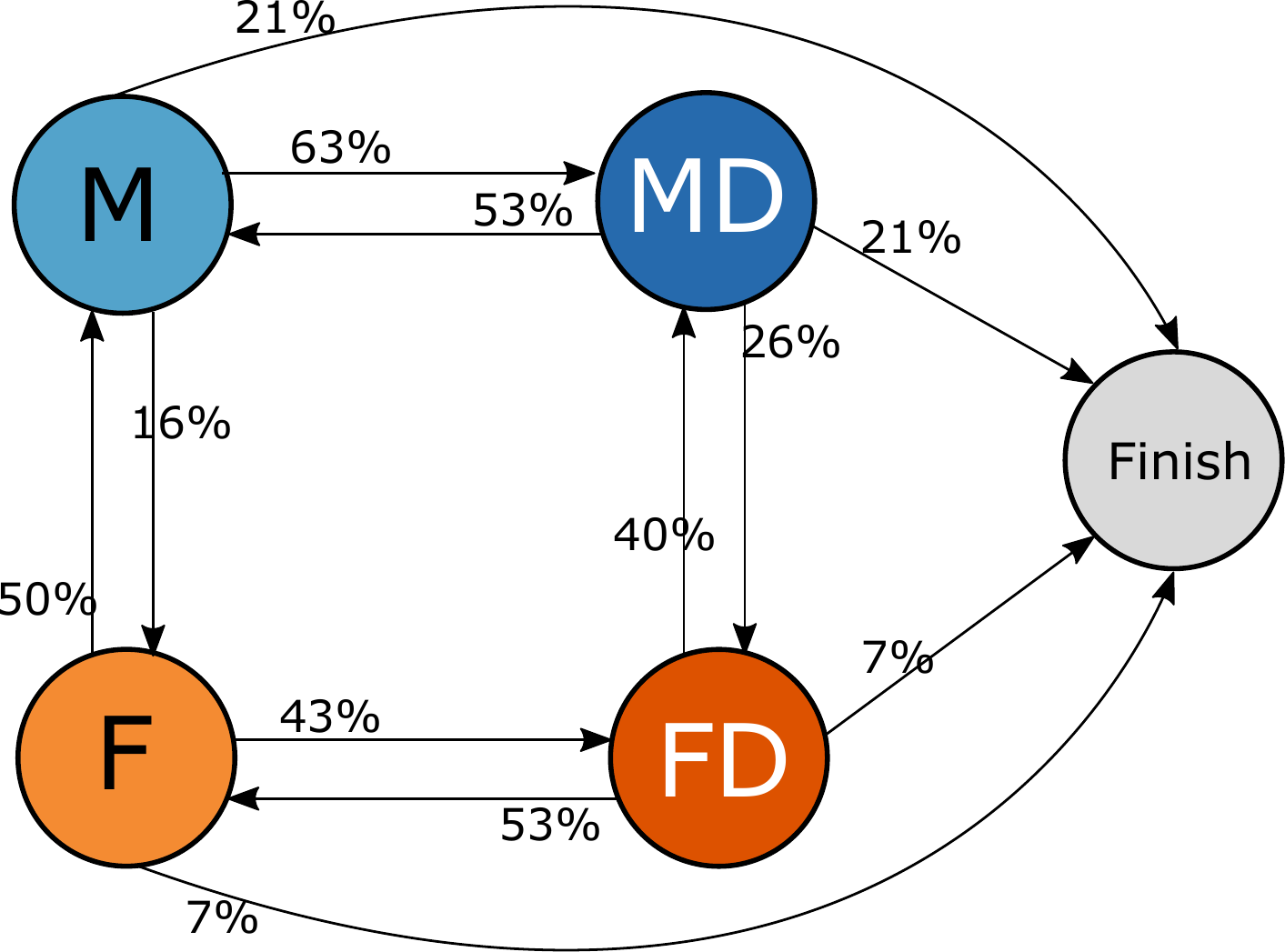}\label{figure:transitions_r5}}
    \hfill
    \subfloat[Task R10 (\emph{complex geometries})]{\includegraphics[width=0.5\columnwidth]{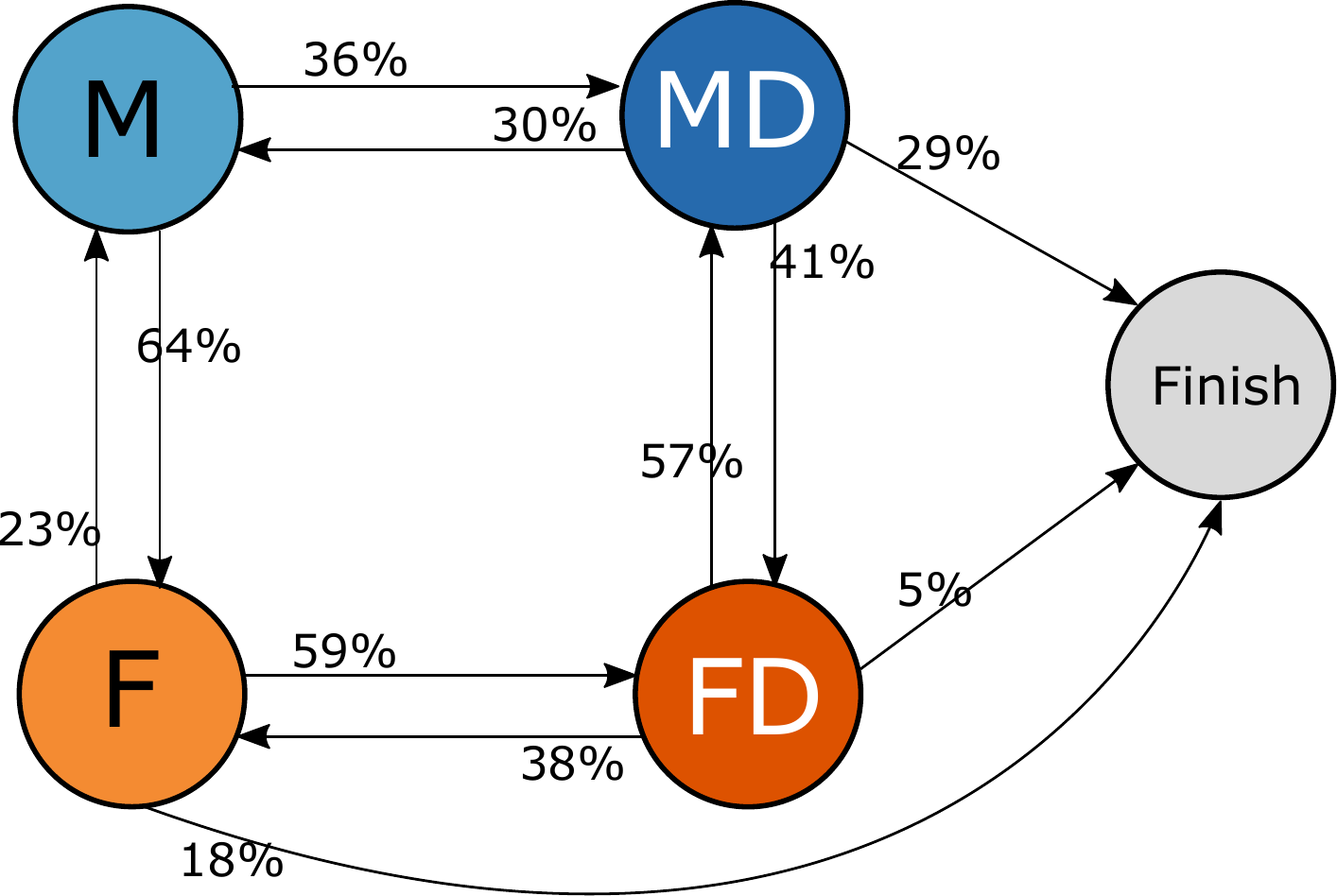}\label{figure:transitions_r10}}
  \caption{Effective interface transition probabilities for Tasks R1, R3, R5, and R10, one pertaining to each of the four categories or use case scenarios (editing surfaces, editing in free space, handling occlusions, and editing complex geometries). Self-loop transitions have been removed.}
\label{figure:interface_transitions}
\end{figure*}

\old{Tasks R4 and R10, which involve editing complex geometries, clearly show the need for the color and depth selection tools. When intricate geometries are present, these are extensively used. The nature of the scene determines which one is used: in the case of R4, 9 out of 10 subjects used the depth selection tool to complete the task, while in the case of R10, 9 out of 10 used the color selection tool, as shown in the sample editing workflow for Task R10 shown in Figure~\ref{figure:workflows}. The majority of the subjects used depth information throughout the tasks; however, differences between the time of use of FD and MD were not significant (Figure~\ref{figure:distribution_R10}).}
\\


\textbf{Effective Interface Transitions:} We analyze the probability of transitioning from one interface to another without considering self-loop transitions. Note that there are different user controls for setting Multiview or Focus, and depth or no depth. Therefore, it is not possible to transition directly from M to FD and vice versa, or from F to MD and vice versa. 
Figure~\ref{figure:interface_transitions} shows the interface transition probabilities for the four representative tasks from Experiment 2: R1, R3, R5, and R10.

In Task R1 (Figure~\ref{figure:transitions_r1}), toggling between MD and FD is very common. This points out that users prefer the use of depth, and change between Multiview and Focus to find the most suitable interface to perform their edits or to check their results. In Task R3 (Figure~\ref{figure:transitions_r3}) users always finish their editing in Multiview (with or without depth). This indicates that they prefer to use Multiview to verify that they placed the object correctly in free space, which concurs with our previous observations. For Tasks R5 (Figure~\ref{figure:transitions_r5}) and R10 (Figure~\ref{figure:transitions_r10}), which involve dealing with occlusions and complex geometries, the new depth and color selection tools determine the workflow, decreasing the importance of the choice of interface. This results in participants switching among interfaces looking for what feels more comfortable for drawing and checking results without any clear patterns on the choice of interface.


\old{In summary, the second experiment confirms the findings of Experiment 1 in most aspects, with a clear exception in occlusion handling and intricate geometries, which are now easily dealt with thanks to the new tools. We also observe that depth information is almost always required, while the differences between the interfaces (Multiview and Focus) become less significant.}
Still, the Multiview paradigm (with or without depth) is the prevailing choice to examine the results, as shown by the fact that the majority of transitions to the Finish state tend to come from this paradigm.
}

\section{Conclusions}
We have tested a set of interfaces and tools for light field editing, and shown that they allow users to satisfactorily edit light fields and perform common, everyday tasks, such as those in Experiment 2, on real, captured, light fields. While previous work has focused on the feasibility of using the light field editing interfaces, tools, and depth reconstruction methods, we present here the main findings in terms of workflow and preferences for the different scenarios that a user may encounter (free space, planar surfaces, occlusions, etc.). In order to do so, we have performed state sequence analysis and hidden Markov chain analysis focusing on both the editing tools and interaction paradigms.

\new{With this new analysis, we have noticed that users quickly understand the high dimensionality of light field images and work on a constant iteration of drawing/erasing and checking the results by navigating across light field views or adjusting the depth. We have also demonstrated that having depth information available allows faster and more accurate editing and is preferred by users. When using that depth information, placing edits at the desired depth is straightforward, which reduces the need to use tools that select views or depth, and narrows their use to result checking. Another key finding is that users generally switch to the Multiview paradigm (with or without depth) to review their work before finishing.} 

Although users' actions and preferences in this study may be partially affected by the particular design of the light field editing tools, which may be rapidly updated by the community, we believe this work contributes to the foundations and provides a solid basis for the development of future tools and interfaces, both for researchers and UI designers. \new{Further, we believe these insights can also be used for processing and editing other high dimensional data, such as BRDFs (bidirectional reflectance distribution functions) or BTFs (bidirectional texture functions), for which data-driven editing techniques remain an open problem~\cite{Jarabo2014btf,Serrano2016}.}

\section*{Acknowledgments}
We thank the anonymous reviewers for their insightful comments, and the members of the Graphics and Imaging Lab for their help.
This project has received funding from the European Research Council (ERC) under the European Union’s Horizon 2020 research and innovation programme (CHAMELEON project, grant agreement No 682080), as well as the Spanish Ministerio de Econom\'ia y Competitividad (TIN2016-78753-P and TIN2016-79710-P). 

\bibliographystyle{IEEEtran}
\bibliography{IEEEabrv,LightFields_references}

\end{document}